\documentclass[twocolumn,aps,pra,showpacs,superscriptaddress,
tightenlines,lengthcheck,sort&compress]{revtex4-1}

\usepackage{multirow}
\usepackage{tabularx}
\usepackage{amssymb}
\usepackage{array}
\usepackage{mathrsfs}
\usepackage{amsmath}
\usepackage[bookmarks=false,
 breaklinks=false,pdfborder={0 0 1},colorlinks=false]
 {hyperref}
\hypersetup{
    colorlinks=true,
    linkcolor=blue,
    citecolor=blue,
    urlcolor=blue
}

\makeatletter
\usepackage{diagbox}
\usepackage{amsfonts}
\usepackage{graphicx}
\usepackage{color}

\makeatother

\begin{document}

\title{Phase-tunable remote nonreciprocal charging in waveguide QED}

\author{Meixi Guo}
\affiliation{Department of Applied Mathematics, The Hong Kong Polytechnic University,
Kowloon 999077, Hong Kong, China}

\author{Jian Huang}
\affiliation{Departamento de F\'{\i}s\'{\i}ca Te\'{o}rica de la Materia Condensada and Condensed Matter Physics Center (IFIMAC), Universidad Aut\'{o}noma de Madrid, E28049 Madrid, Spain}

\author{Rui-Yang Gong}
\affiliation{School of Physics, Sun Yat-sen University, Guangzhou 510275, China}

\author{Xian-Li Yin}
\email{xianliyin@foxmail.com}
\affiliation{Department of Applied Mathematics, The Hong Kong Polytechnic University,
Kowloon 999077, Hong Kong, China}

\author{Guofeng Zhang}
\email{guofeng.zhang@polyu.edu.hk}
\affiliation{Department of Applied Mathematics, The Hong Kong Polytechnic University,
Kowloon 999077, Hong Kong, China and The Hong Kong Polytechnic University
Shenzhen Research Institute, Shenzhen, Guang Dong 518057, China}
\affiliation{Research Institute for Quantum Technology, The Hong Kong Polytechnic
University, Hong Kong, China}

\begin{abstract}
Remote quantum batteries require directional and controllable energy transfer between spatially separated quantum nodes, yet most existing protocols rely on direct charger-battery Hamiltonian couplings. Here we propose a phase-tunable waveguide-QED architecture for remote quantum-battery charging, in which a driven charger and a remote battery are coupled solely via engineered waveguide-mediated interference, without any direct local interaction. We systematically compare four configurations: two-giant-emitter and giant-small-emitter hybrids, each with open or mirror-terminated waveguides. By engineering the propagation and coupling phases, the waveguide-mediated coherent exchange interaction and collective dissipation can be balanced to suppress the backward channel while retaining a finite forward channel, thereby realizing cascaded-like unidirectional charging. Our analysis shows that nonreciprocity and storage efficiency can be independently engineered, offering design flexibility for different quantum network scenarios. The giant-small-emitter mirror-terminated configuration simultaneously achieves perfect nonreciprocity and battery-dominated storage, while both giant-small-emitter configurations exhibit distance-insensitive directionality. Extending the scheme to quadratic driving, we show that anomalous second moments render the battery state non-passive, making ergotropy a performance metric distinct from stored energy. These results establish phase-tunable waveguide networks as a versatile platform for remote quantum-energy transfer and provide design principles for directional and work-extractable energy storage in quantum networks.
\end{abstract}

\date{\today }

\maketitle

\section{Introduction}

Thermodynamics provides a universal framework for classical energy machines, while the continued miniaturization of energy-storage and information-processing devices has made genuinely quantum effects relevant for energy conversion and storage, motivating the development of quantum thermodynamics~\cite{KosloffLevy2014, FCampaioli2024Colloquium,uzdin2016quantumthermo}.
Against this backdrop, the concept of a quantum battery (QB), which refers to a quantum system that stores energy and releases it on demand, was introduced~\cite{alicki2013entanglement,FCampaioli2024Colloquium}.
Recent theoretical studies have explored how quantum coherence and correlations~\cite{shi2022entanglement,caravelli2021energy,andolina2018collective}, 
collective many-body effects~\cite{binder2015quantacell,campaioli2017enhancing,ferraro2018high,le2018spin,rossini2020syk}, 
and optimized charging protocols and power bounds~\cite{garciapintos2020fluctuations,FCampaioli2024Colloquium} 
can enhance charging performance.
Experimental proof-of-principle demonstrations have also begun to appear in organic microcavities~\cite{quach2022scadv}, 
superconducting circuits~\cite{hu2022qst}, 
and nuclear-spin platforms~\cite{joshi2022pra}, 
along with controlled energetic characterization of energy transfer between quantum emitters and light fields~\cite{wenniger2023energytransfer}.

Many QB protocols rely on an engineered local Hamiltonian coupling between the charger and the battery~\cite{andolina2018collective,binder2015quantacell,le2018spin}. This local-coupling paradigm provides direct control over the charging rate and offers a natural starting point for studying fundamental charging mechanisms and performance bounds. In many practical architectures, however, the charger and battery are spatially separated. Direct interactions are typically short-ranged and present significant challenges in implementation, tuning, and stabilization within scalable configurations~\cite{song2024remote}. 
These limitations motivate environment-mediated charging strategies, where a common reservoir mediates interactions between spatially separated nodes~\cite{tabesh2020environment,xu2021environment,song2024remote}. Within this framework, a common reservoir can generate charger--battery correlations even in the absence of direct coupling, enabling finite steady-state energy storage~\cite{kamin2024nonmarkovian}, collective charging enhancement~\cite{carrasco2022collective}, self-discharging suppression~\cite{song2024remote}, and high-power energy stabilization~\cite{quach2020dark,yao2021stable} in environment-mediated quantum batteries.

To go beyond the simple environment-mediated charging effects described above and achieve directional, long-range energy transfer, reservoir engineering provides a powerful tool~\cite{poyatos1996reservoir,wang2023nonreciprocal,yang2026quantum}. By balancing coherent exchange interaction against collective dissipation, one can cancel the effective coupling in one direction while retaining it in the opposite direction, thereby realizing unidirectional energy transport~\cite{Metelmann2015}. Such nonreciprocal coupling not only ensures that energy flows exclusively from the charger to the battery, but also protects the battery from backflow that would otherwise reduce the net stored energy. Moreover, the directionality can be switched by reconfiguring the interference conditions, offering flexibility for different charging protocols. This directional control has recently been exploited to enhance charging performance in nonreciprocal quantum battery proposals~\cite{ahmadi2024nonreciprocal,ahmadi2025applied,guo2025nonreciprocal,sun2025nonreciprocal,lin2026enhanced,xu2025enhanced}.

Despite these advances, a systematic comparison of different coupling configurations and their impact on both directionality and storage efficiency remains absent. 
Waveguide QED provides an ideal platform to address this gap, thanks to its flexible coupling architecture, one-dimensional propagation channel, and phase-tunable interference~\cite{liao2016photon,roy2017colloquium,gu2017microwave,sheremet2023waveguide}. 
Hence, it enables collective charging in extended waveguide-QED battery architectures through collective effects~\cite{tirone2025many,yin2025feedback}, and also supports remote energy transfer by confining the electromagnetic field to a one-dimensional channel~\cite{song2024remote,lu2025topological,liu2026chiral}. 
By engineering photon-mediated coherent coupling and collective dissipation, one can therefore control the directionality and thus enhance the efficiency of long-distance charging.
Beyond the small-emitter limit, giant emitters~\cite{kockum2020quantum} have attracted considerable attention in recent years as a versatile platform for waveguide QED. 
Because giant atoms couple to a waveguide at multiple spatially separated points, the usual dipole approximation breaks down~\cite{walls2008quantum} and quantum self-interference becomes an intrinsic resource. This multi-point structure gives rise to frequency-dependent waveguide coupling rates and Lamb shifts~\cite{frisk2014designing}, decoherence-free interactions~\cite{kannan2020waveguide,kockum2018decoherence,carollo2020mechanism,soro2022chiral,du2023complex}, bound states~\cite{guo2020oscillating,wang2021tunable,cheng2022topology,qiu2023collective}, and entanglement generation~\cite{du2025dressed,santos2023generation,yin2023generation,yin2025giant}.
These phenomena originate primarily from the propagation phases acquired by photons traveling between coupling points. Furthermore, introducing time modulation to the emitter-waveguide couplings induces additional local coupling phases. By tuning both propagation and coupling phases, one can achieve chiral emission and more complex quantum interference pathways~\cite{wang2022chiralnetwork,joshi2023resonance,chen2022nonreciprocal,chang2025nonmarkovian}. 
Recent work has also begun to explore giant-atom quantum batteries and interference-engineered energy transfer~\cite{yan2026giantatomQB}. 
Nevertheless, how to realize directional, nonreciprocal energy transfer over long distances in the simultaneous presence of external driving and dissipation remains an open and interesting problem.

In this paper, we propose and analyze a waveguide-mediated remote quantum battery where a driven charger and a spatially separated battery, both modeled as bosonic modes, are coupled solely through engineered waveguide interference, without any direct local interaction. We systematically compare four coupling geometries: open and mirror-terminated versions of two-giant-emitter and giant-small-emitter hybrids. By tuning the propagation and coupling phases, we establish a nonreciprocal working point that suppresses backward energy flow while preserving forward transfer, thereby realizing cascaded-like unidirectional charging. For linear coherent driving, we introduce two figures of merit: the nonreciprocal ratio quantifies directional asymmetry, and the relative storage ratio determines whether the battery stores more energy than the charger. Our analysis reveals that perfect nonreciprocity and battery-dominated storage are governed by distinct interference conditions and can therefore be engineered independently. The mirror-terminated giant-small-emitter coupling simultaneously achieves both objectives, while the open giant-small-emitter coupling configuration maintains robust directionality with negligible sensitivity to the distance between the charger and the battery. We further extend the analysis to quadratic driving~\cite{zhang2019powerful,crescente2020charging,crescente2020ultrafast,downing2023quantum,xu2025enhanced}, where the same phase-engineered couplings support stable directional charging. However, the quadratic drive generates anomalous second moments that induce squeezing, making the battery state non-passive, and ergotropy becomes a performance metric distinct from stored energy~\cite{leghtas2015confining,wang2019parametric,savona2017phase,rota2019critical,barra2022phase}. Among the four configurations, the mirror-terminated giant-small-emitter setup again yields the largest extractable work fraction and the widest parameter region where the battery stores more energy than the charger.

The remainder of the paper is organized as follows. In Sec.~\ref{sec:theory}, we introduce the waveguide-mediated model, derive the forward and backward directional couplings, and define the relevant performance indicators. In Sec.~\ref{sec:linear}, we analyze remote nonreciprocal charging under linear driving and compare the four configurations in terms of directionality and storage efficiency. In Sec.~\ref{subsec:Quadratic}, we extend the analysis to quadratic driving, discussing stability, anomalous second moments, and ergotropy. Finally, the results are concluded in Sec.~\ref{sec:conclusion_outlook}.


\begin{figure}[t]
    \centering
    \includegraphics[width=\linewidth]{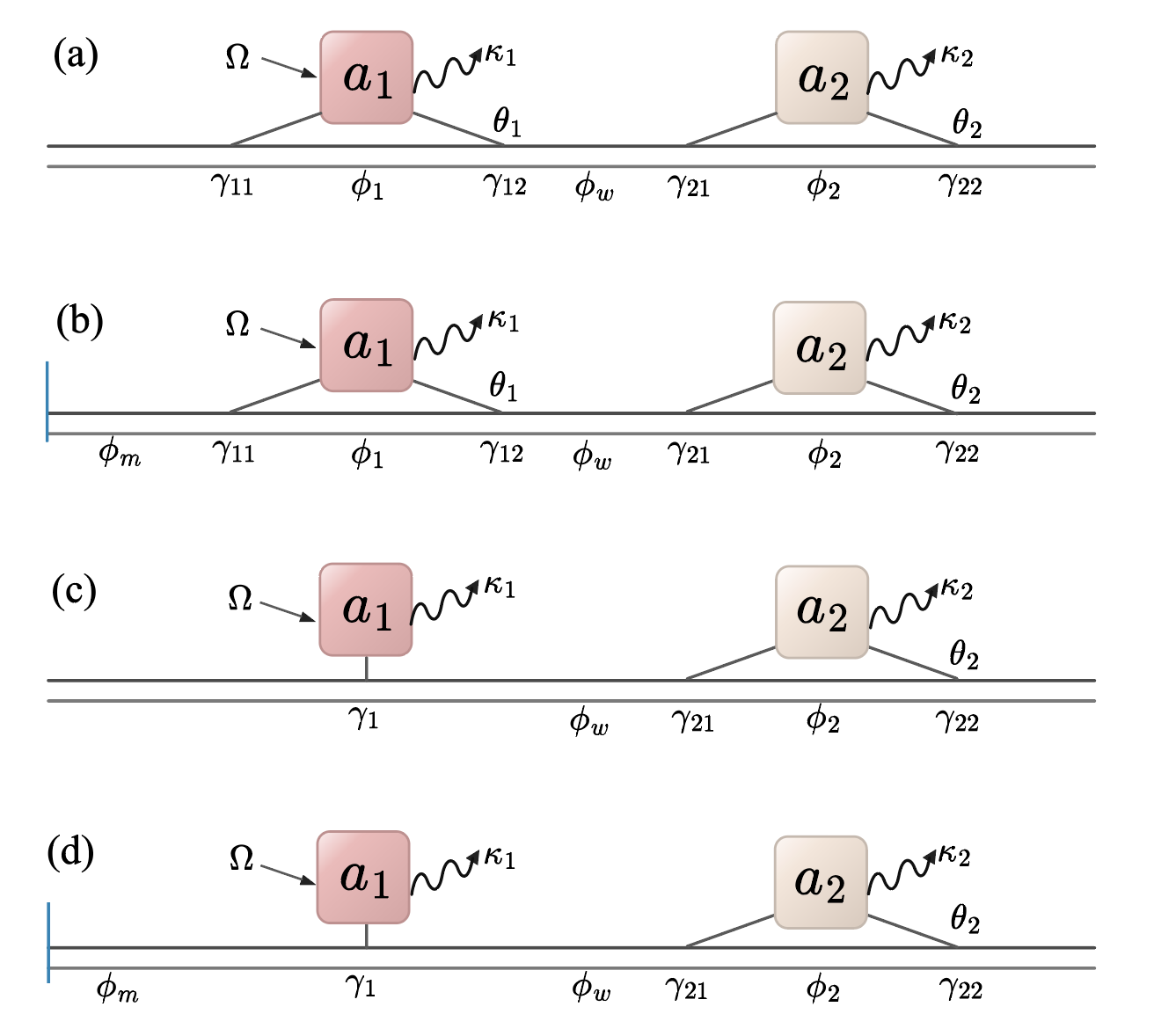}
    \caption{
Schematic illustration of four different waveguide-mediated nonreciprocal QB configurations:
(a) Open four-point model (4P),   
(b) Mirror-terminated four-point model (4P+M),   
(c) Open three-point model (3P), and
(d) Mirror-terminated three-point model (3P+M). 
Each setup consists of two single-mode cavities (a driven charger and a remote QB) coupled to a one-dimensional waveguide. In the four-point designs (a,b), both cavity modes are coupled to the waveguide at two spatially separated points with waveguide coupling rates $\gamma_{11}$, $\gamma_{12}$, $\gamma_{21}$, and $\gamma_{22}$. In the three-point designs (c,d), the left cavity mode is configured as a small emitter and couples to the waveguide at a single point with coupling rate $\gamma_{1}$, while the right cavity mode is configured as a two-point giant emitter with coupling rates $\gamma_{21}$ and $\gamma_{22}$. The propagation phases $\phi_1$, $\phi_w$, $\phi_2$ and the mirror phase $\phi_m$ (for the mirror-terminated case), control the interference pathways. The local phases $\theta_1$ and $\theta_2$ denote the relative coupling phases associated with the two coupling points of the left and right giant-emitter modes, respectively; in the three-point geometries only $\theta_2$ is present. In superconducting-circuit implementations, such phase-dependent multipoint interference can be engineered using giant artificial atoms and tunable waveguide couplings~\cite{joshi2023resonance,kockum2018decoherence,wang2021tunable}. The external driving amplitude is $\Omega$. $\kappa_1$ and $\kappa_2$ denote the intrinsic cavity loss rates. }
    \label{fig:schematic4p}
\end{figure}

\section{Theoretical model}
\label{sec:theory}

\subsection{Quantum master equation for the charger--battery systems}
\label{subsec:master_equation}

As illustrated in Fig.~\ref{fig:schematic4p}, we consider four types of waveguide-QED configurations for remote charging of a  quantum battery. In each setup, two spatially separated single-mode cavities are coupled to a one-dimensional waveguide. When one cavity mode is driven by an external pump field, it acts as the charger, while the other serves as the QB. The external drive can be applied to either the left or the right cavity mode, thereby interchanging the operational roles of the two modes. In all setups, we assume that the two cavity modes are initially in their vacuum state and the waveguide is in the vacuum state.

In Figs.~\ref{fig:schematic4p}(a) and \ref{fig:schematic4p}(c) the waveguide is open, whereas in Figs.~\ref{fig:schematic4p}(b) and \ref{fig:schematic4p}(d) it is semi-infinite with a mirror terminating the left end. In Figs.~\ref{fig:schematic4p}(a) and \ref{fig:schematic4p}(b), both
cavities are giant emitters, each coupled to the waveguide at two separate coupling points. In Figs.~\ref{fig:schematic4p}(c) and \ref{fig:schematic4p}(d), the left cavity mode $a_1$ is configured as a small emitter and couples to the waveguide at a single point, while the right cavity mode $a_2$ remains a two-point giant emitter. For convenience, we refer to these four configurations as 4P (four-point, open), 4P+M (four-point, mirror-terminated), 3P (three-point, open), and 3P+M (three-point, mirror-terminated), respectively.

To describe the charging dynamics, we derive a quantum master equation for the density matrix $\rho$ of the charger and QB by tracing out the continuum field modes in the waveguide. To this end, we assume that the coupling between each cavity mode and the waveguide field is weak compared to the bare resonance frequencies of the modes. In addition, the photon propagation time between coupling points is assumed to be much shorter than the characteristic system timescales and retardation effects can be neglected. Using the SLH formalism~\cite{GoughJames2009CMP,GoughJames2009IEEE,CombesKerckhoffSarovar2017}, the quantum master equation for all setups in Fig.~\ref{fig:schematic4p} can be written as ($\hbar=1$)

\begin{equation}
\label{eq:master}
\begin{split}
\dot{\rho}
= {} & -i\Big[
\Delta_{1}^{\mathrm{eff}}a_{1}^{\dagger}a_{1}
+\Delta_{2}^{\mathrm{eff}}a_{2}^{\dagger}a_{2}
+\big(Ja_{2}^{\dagger}a_{1}+\mathrm{H.c.}\big)
+H_{d},
\rho\Big] \\
& +\Lambda_1\mathcal{D}[a_1]\rho
  +\Lambda_2\mathcal{D}[a_2]\rho \\
& +\Big[\Gamma_{12}\!\left(a_1\rho a_{2}^{\dagger}
-\tfrac{1}{2}\{a_{2}^{\dagger}a_1,\rho\}\right)
+\mathrm{H.c.}\Big],
\end{split}
\end{equation}
where we have moved to a frame rotating at the driving frequency $\omega_L$.
Here, $a_1$ ($a_1^\dagger$) and $a_2$ ($a_2^\dagger$) are the annihilation (creation) operators of the two cavity modes, respectively.
The effective detunings are defined as
$\Delta_{1}^{\mathrm{eff}}=\Delta_{1}+\delta\omega_{1}$ and
$\Delta_{2}^{\mathrm{eff}}=\Delta_{2}+\delta\omega_{2}$, where
$\Delta_{1}=\omega_{1}-\omega_L$ and
$\Delta_{2}=\omega_{2}-\omega_L$ are the detunings between the cavity-mode resonance frequencies and the driving frequency, and
$\delta\omega_1$ ($\delta\omega_2$) is the waveguide-induced Lamb shift of cavity mode $a_1$ ($a_2$).
The coefficient $J$ denotes the waveguide-mediated coherent exchange interaction between the charger and the QB.
The total linewidths are
$\Lambda_1=\Gamma_1+\kappa_1$ and
$\Lambda_2=\Gamma_2+\kappa_2$, where
$\Gamma_1$ and $\Gamma_2$ are the waveguide-induced individual decay rates of cavity modes $a_1$ and $a_2$, respectively, while
$\kappa_1$ and $\kappa_2$ represent their intrinsic cavity loss rates.
Finally, $\Gamma_{12}$ is the waveguide-mediated collective decay rate between the two cavity modes.
The explicit expressions of $\delta\omega_1$, $\delta\omega_2$, $J$, $\Gamma_1$, $\Gamma_2$, and $\Gamma_{12}$ for the four configurations are collected in Appendix~\ref{app:coefficients}, Table~\ref{tab:coeff_all_4col}.

The driving Hamiltonian $H_d$ describes the external pumping applied to the system. In our nonreciprocal charging scheme, we consider two different types of driving: linear driving and quadratic driving.  For the left-mode-driven case, the external drive is applied to cavity mode $a_1$. For linear driving, the drive Hamiltonian reads
\begin{equation}
H_d = H_d^{(\mathrm{lin})}=\Omega\,(a_{1}^{\dagger}+a_{1}),
\label{eq:Hd_lin}
\end{equation}
where $\Omega$ is the driving amplitude. For quadratic driving, the driving Hamiltonian is given by 
\begin{equation}
H_d = H_d^{(\mathrm{quad})}=\frac{\Omega}{2}(a_{1}^{\dagger 2}+a_{1}^{2}).
\label{eq:Hd_quad}
\end{equation} 
The right-mode-driven case is obtained by simply replacing $a_1$ with $a_2$ in Eqs.~\eqref{eq:Hd_lin} and \eqref{eq:Hd_quad}, while the other terms in quantum master equation Eq.~\eqref{eq:master} remain unchanged regardless of which mode is driven. These two driving schemes are chosen to contrast fundamentally different charging mechanisms. In the following, we investigate in detail the  nonreciprocal charging dynamics and QB performance under these two driving schemes.

\subsection{Performance indicators of the QB}

In order to characterize the charging performance of the QB, we focus on two quantities: the stored energy $E$ and the corresponding ergotropy $\mathcal{W}$~\cite{Allahverdyan2004,Francica2020}, which quantify the QB's stored energy and maximum extractable work, respectively. We assume that both the charger and QB are initialized in the vacuum states. The energy stored in the QB is defined as
\begin{equation}
E_b(t) = \operatorname{Tr}\!\left( H_b\,\rho_b\right)
=\omega_b\langle a_b^\dagger a_b\rangle,
\label{eq:Eb_def}
\end{equation}
where $a_b$ denotes the remote battery mode for the chosen driving direction: $a_b=a_2$ under left-mode driving and $a_b=a_1$ under right-mode driving. Correspondingly, $\omega_b=\omega_2$ for left-mode driving and $\omega_b=\omega_1$ for right-mode driving. Here, $H_b=\omega_b a_b^\dagger a_b$ is the Hamiltonian of the QB, $\rho_b(t)=\operatorname{Tr}_c[\rho(t)]$ is the reduced state of the battery.

The ergotropy $\mathcal{W}_b$ of the QB, which quantifies the maximal extractable work, can be obtained from the QB's state $\rho_b$ by cyclic unitary operations (with $H_b$ fixed). It can be expressed as the difference between the stored energy and the passive-state energy~\cite{alicki2013entanglement, Allahverdyan2004},
\begin{equation}
\mathcal{W}_b(t) = E_b(t)-E_b^\beta(t).
\label{eq:ergotropy_def}
\end{equation}
Here $E_b^\beta(t)$ is the energy of the passive state associated with $\rho_b$, i.e., the part of the stored energy that cannot be extracted by any cyclic unitary operation. It can be expressed by~\cite{alicki2013entanglement} 
\begin{equation}
E_b^\beta(t) = \operatorname{Tr}( H_b\,\rho_b^\beta)
=\omega_b \frac{\sqrt{D_b(t)}-1}{2},
\label{eq:Eb_beta}
\end{equation}
where we introduce the dimensionless parameter $D_b$, which is given by 
\begin{equation}
D_b(t)=(1+2\langle a_b^\dagger a_b\rangle-2|\langle a_b\rangle|^2)^2
-4|\langle a_b a_b\rangle-\langle a_b\rangle^2|^2.
\label{eq:Db_def}
\end{equation}
Equation~\eqref{eq:Db_def} indicates that  $D_b(t)$ is determined by the first and second moments of the battery mode. This quantity is the determinant of the single-mode covariance matrix and therefore determines the symplectic eigenvalue, or equivalently the thermal occupation of the associated passive state.

\section{Remote nonreciprocal charging in the linear-driving case}\label{sec:linear}
In this section, we compare remote charging for the four different coupling  configurations in Fig.~\ref{fig:schematic4p}. Here, we consider that the charger is supplied energy by a linear coherent driving, with the Hamiltonian $H_d^{(\mathrm{lin})}$ taking the form given in Eq.~\eqref{eq:Hd_lin}. By using Eqs.~\eqref{eq:master} and \eqref{eq:Hd_lin}, the time evolution of the first moments can be obtained as
\begin{subequations}\label{eq:EoMset}
\begin{align}
\dot{\langle a_1\rangle}
&= -\left(i\Delta_{1}^{\mathrm{eff}}+\frac{\Lambda_{1}}{2}\right)\langle a_1\rangle
-\left(iJ^{*}+\frac{\Gamma_{12}^{*}}{2}\right)\langle a_2\rangle
   - i\Omega,
\label{eq:set_a}\\[2mm]
\dot{\langle a_2\rangle}
&= -\left(i\Delta_{2}^{\mathrm{eff}}+\frac{\Lambda_{2}}{2}\right)\langle a_2\rangle
   -\left(iJ+\frac{\Gamma_{12}}{2}\right)\langle a_1\rangle.
\label{eq:set_b}
\end{align}
\end{subequations}
The right-mode-driven case is obtained by interchanging the roles of $a_1$ and $a_2$ in the driving term. Equation~\eqref{eq:EoMset} indicates that, for left-mode driving, the coupling between the charger $a_1$ and the QB $a_2$ is governed by two effective coefficients,
\begin{equation}
g_{\rightarrow}=-iJ-\frac{\Gamma_{12}}{2},
\qquad
g_{\leftarrow}=-iJ^{*}-\frac{\Gamma_{12}^{*}}{2}.
\label{eq:g}
\end{equation}
Here, $g_{\rightarrow}$ characterizes the forward influence from the charger $a_1$ to the QB $a_2$, while $g_{\leftarrow}$ describes the backward influence from $a_2$ to $a_1$. Importantly, both the coherent exchange $J$ and the collective decay rates originate purely from waveguide-mediated interference, without any direct local coupling between the charger and the QB. As summarized in Appendix~\ref{app:coefficients} (Table~\ref{tab:coeff_all_4col}), these coefficients can be tuned on demand by engineering the coupling and propagation phases.

To evaluate the energy supplied to the charger, as well as the stored energy and maximal extractable energy of the QB, we calculate the time evolution of the relevant second-order moments. Based on the quantum master equation Eq.~\eqref{eq:master} and the linear driving Hamiltonian Eq.~\eqref{eq:Hd_lin}, we obtain a closed system of evolution equations for these second moments. For the left-mode-driven case,
\begin{subequations}\label{eq:EoMsecond}
\begin{align}
\dot{\langle a_1^\dagger a_1\rangle}
&= -\Lambda_1\langle a_1^\dagger a_1\rangle
+g_{\leftarrow}^{*}\langle a_2^\dagger a_1\rangle
+g_{\leftarrow}\langle a_1^\dagger a_2\rangle
\nonumber\\
&\quad + i\Omega\big(\langle a_1\rangle-\langle a_1^\dagger\rangle\big),
\label{eq:set_d}\\[1mm]
\dot{\langle a_2^\dagger a_2\rangle}
&= -\Lambda_2\langle a_2^\dagger a_2\rangle
+g_{\rightarrow}^{*}\langle a_1^\dagger a_2\rangle
+g_{\rightarrow}\langle a_2^\dagger a_1\rangle,
\label{eq:set_c}\\
\dot{\langle a_1^\dagger a_2\rangle}
&=\left[-i\!\left(\Delta_2^{\mathrm{eff}}-\Delta_1^{\mathrm{eff}}\right)
-\frac{\Lambda_1+\Lambda_2}{2}\right]\langle a_1^\dagger a_2\rangle
\nonumber\\
&\quad +g_{\leftarrow}^{*}\langle a_2^\dagger a_2\rangle 
+g_{\rightarrow}\langle a_1^\dagger a_1\rangle
+ i\Omega\,\langle a_2\rangle,
\label{eq:set_f}\\
\dot{\langle a_2^\dagger a_1\rangle}
&=\left[i\!\left(\Delta_2^{\mathrm{eff}}-\Delta_1^{\mathrm{eff}} \right)
-\frac{\Lambda_1+\Lambda_2}{2}\right]\langle a_2^\dagger a_1\rangle
\nonumber\\
&\quad +g_{\rightarrow}^{*}\langle a_1^\dagger a_1\rangle + g_{\leftarrow}\langle a_2^\dagger a_2\rangle
- i\Omega\,\langle a_2^\dagger\rangle.
\label{eq:set_e}
\end{align}
\end{subequations}
In principle, the full set of moment equations can be solved exactly. Under the linear coherent driving and starting from the global vacuum, however, the field states remain coherent, and the normally ordered moments factorize as products of the first-order moments~\cite{Farina2019},
\begin{equation}
\langle xy\rangle_{\mathrm{co}}=\langle x\rangle_{\mathrm{co}}\langle y\rangle_{\mathrm{co}},
\qquad x,y\in\{a_1,a_2,a_1^\dagger,a_2^\dagger\}.
\label{eq:factorization_coherent}
\end{equation}

According to Eq.~\eqref{eq:factorization_coherent}, we have $\langle a_2^\dagger a_2\rangle = |\langle a_2\rangle|^2$ and $\langle a_2 a_2\rangle=\langle a_2\rangle^2$ (and analogously for $a_1$). Inserting these two simplifications into Eq.~\eqref{eq:Db_def}, we have $D_b(t)=1$ at all times, and the passive state energy $E_b^\beta(t)=0$. In this case, the stored energy in the QB is fully extractable. Therefore, we only focus on the stored energy of the QB in the coherent-driving case. We set the effective detunings $\Delta_1^{\mathrm{eff}}=\Delta_2^{\mathrm{eff}}=0$, which requires $\Delta_i=\omega_i-\omega_L=-\delta\omega_i$ with $i=1,2$. Since the Lamb shifts $\delta\omega_1$ and $\delta\omega_2$ may be different (see Table~\ref{tab:coeff_all_4col}), the corresponding detunings $\Delta_1$ and $\Delta_2$ can differ as well, which generally leads to $\omega_1 \neq \omega_2$. However, the frequency mismatch is typically small (on the order of the waveguide coupling rates $\gamma_j$). Solving Eqs.~\eqref{eq:EoMset} and \eqref{eq:EoMsecond}, the stored energy in the QB for the left-mode-driven case is given by
\begin{equation}
\label{eq:nb_t_t_general}
\begin{aligned}
E_b^{\mathrm{L}}(t)
&= \frac{256 \,\omega_2 \,|g_{\rightarrow}|^{2}\Omega^{2}\,e^{-\frac{1}{2}\Lambda t}}
{\left|S^{2}-\Lambda^{2}\right|^{2}} \\
&\quad\times
\left|
\cosh\!\left(\frac{St}{4}\right)
+\frac{\Lambda}{S}\sinh\!\left(\frac{St}{4}\right)
-e^{\frac{\Lambda t}{4}}
\right|^{2},
\end{aligned}
\end{equation}
where $\Lambda=\Lambda_1+\Lambda_2$, $\delta=\Lambda_2-\Lambda_1$ and $S=\sqrt{\delta^{2}+16\,g_{\leftarrow} g_{\rightarrow}}$. We can also obtain the stored energy $E_c^\mathrm{L}(t)=\omega_1 \langle a_1^\dagger a_1\rangle$ in the charger as

\begin{eqnarray}
\label{eq:nc_t_general}
E_{c}^{\text{L}}(t) &=&\frac{\omega _{1}\Omega ^{2}}{|4g_{\leftarrow}g_{\rightarrow }-\Lambda _{1}\Lambda _{2}|^{2}}  \nonumber \\
&&\times \left \vert \Lambda +\delta -e^{-\frac{\Lambda t}{4}}\left[ (\Lambda+\delta )\cosh \left( \frac{St}{4}\right) \right. \right.  \\
&&+\left. \left. \frac{S^{2}+\Lambda \delta }{S}\sinh \left( \frac{St}{4}\right) \right] \right \vert ^{2}.  \nonumber
\end{eqnarray}


\begin{table}[t]
\caption{\label{tab:nonreciprocal_points} Phase parameters satisfying the nonreciprocal condition $g_{\leftarrow}=0$.}
\begin{ruledtabular} 
\begin{tabular}{ll} 
\textbf{Setups} & \textbf{Phase choices} \\ 
\hline 
4P & $\phi_1=0,\ \phi_w=\pi,\ \phi_2=\frac{\pi}{2},\ \theta_1=0,\ \theta_2=\frac{\pi}{2}$ \\ 
4P+M & $\phi_m=\frac{2\pi}{3},\ \phi_1=0,\ \phi_w=\frac{\pi}{4},\ \phi_2=\frac{\pi}{6},\ \theta_1=0,\ \theta_2=\frac{3\pi}{2}$ \\ 
3P & $\phi_w=\pi,\ \phi_2=\frac{\pi}{2},\ \theta_2=\frac{\pi}{2}$ \\ 
3P+M & $\phi_m=0,\ \phi_w=\pi,\ \phi_2=\frac{\pi}{2},\ \theta_2=\frac{\pi}{2}$ \\ 
\end{tabular} 
\end{ruledtabular}
\end{table}

To achieve nonreciprocal charging, we can balance the exchange interaction strength $J$ and collective decay rate $\Gamma_{12}$ \cite{Metelmann2015,ahmadi2024nonreciprocal,sun2025nonreciprocal}. Concretely, by engineering the local coupling phases and the propagation phases of the four setups in Fig.~\ref{fig:schematic4p}, one can realize $J^*=i\Gamma_{12}^*/2$, which, upon substitution into Eq.~\eqref{eq:g}, yields $g_{\leftarrow}=0$. Thus, the backward dynamical influence from the QB to the charger is completely suppressed, while the forward coupling remains finite. Under this condition, the energy flows unidirectionally from the charger to the QB without any backflow, realizing a cascaded-like dynamics mediated only by the waveguide. By imposing the condition $g_{\leftarrow}=0$, one obtains the phase relations required for nonreciprocal charging. Table~\ref{tab:nonreciprocal_points} lists representative values of the coupling phases and propagation phases for each configuration, which will be used in later simulations. For the 3P and 3P+M setups, however, once the other phases are fixed, the choice of $\phi_w$ remains free. Under the nonreciprocal condition, Eqs.~\eqref{eq:nb_t_t_general} and \eqref{eq:nc_t_general} reduce to
\begin{subequations}\label{eq:nonreciprocal_energies}
\begin{align}
E_b^{\mathrm{L,nr}}(t)
&=
\frac{16\omega_2 |g_{\rightarrow}|^2\Omega^2}
{\Lambda_1^2\Lambda_2^2}
\nonumber\\
&\quad\times
\left|
1-e^{-\Lambda t/4}
\left[
\cosh\!\left(\frac{\delta t}{4}\right)
+\frac{\Lambda}{\delta}
\sinh\!\left(\frac{\delta t}{4}\right)
\right]
\right|^2,
\label{eq:Eb_nonreciprocal_t}\\
E_c^{\mathrm{L,nr}}(t)
&=
\frac{4\omega_1\Omega^2}{\Lambda_1^2}
\left(1-e^{-\Lambda_1t/2}\right)^2.
\label{eq:Ec_nonreciprocal_t}
\end{align}
\end{subequations}
Under the nonreciprocal condition, the left-driven charger equation in Eq.~\eqref{eq:EoMset} no longer contains the QB amplitude, and the charger therefore reduces to a single driven damped mode with linewidth $\Lambda_1$, as reflected in Eq.~\eqref{eq:Ec_nonreciprocal_t}. The QB dynamics, by contrast, is governed by the remaining forward waveguide-mediated coupling $g_{\rightarrow}$ and by the linewidths $\Lambda_1$ and $\Lambda_2$, which gives the $|g_{\rightarrow}|^2/(\Lambda_1^2\Lambda_2^2)$ scaling in Eq.~\eqref{eq:Eb_nonreciprocal_t}. These expressions show that the nonreciprocal condition realizes directional cascaded charging, while the magnitude of the stored QB energy is further determined by the competition between forward transfer and dissipative loss.
\begin{figure}[t]
    \centering
    \includegraphics[width=\linewidth]{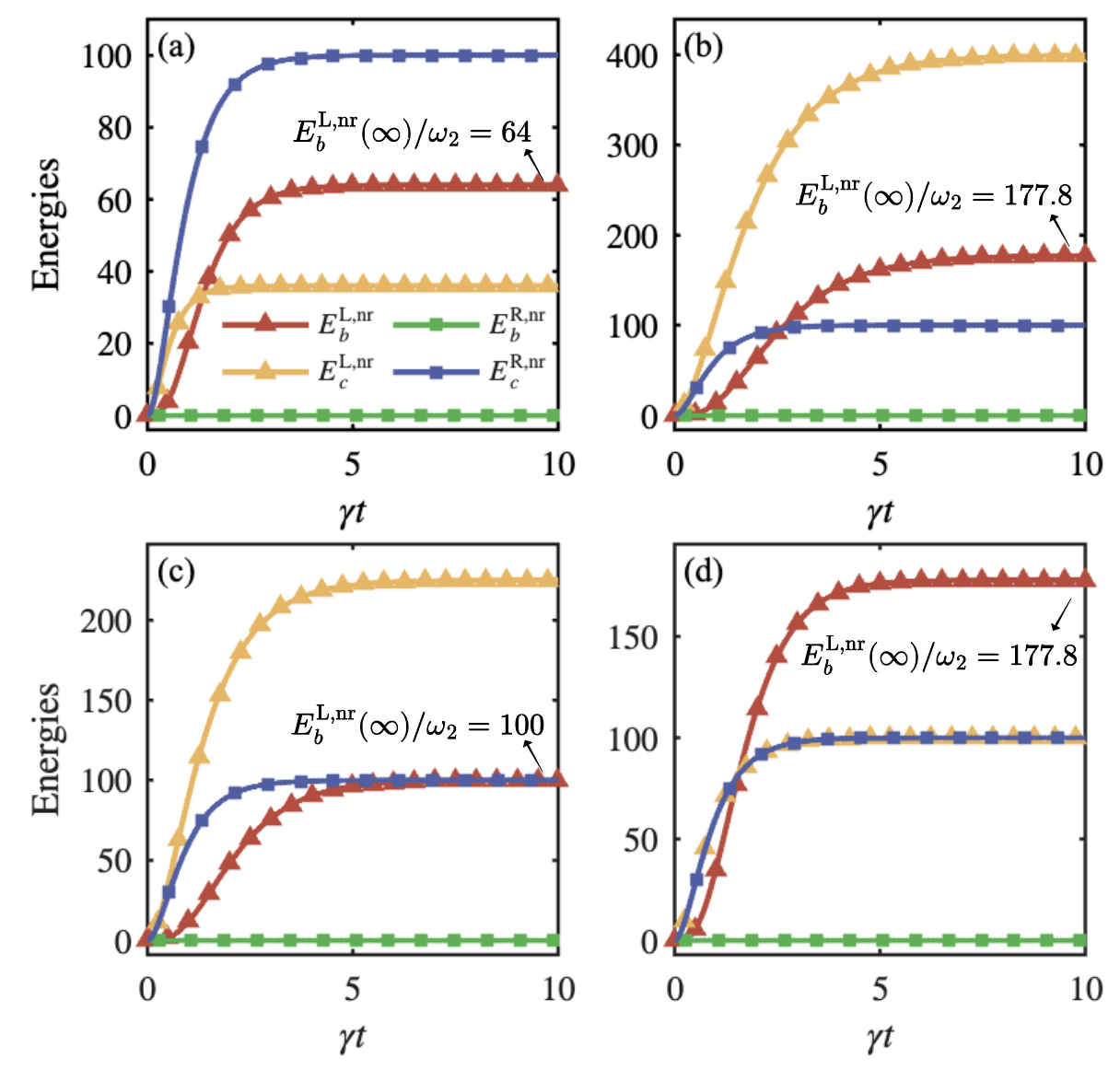}
    \caption{\label{fig:time_dynamics}
Nonreciprocal charging dynamics for different setups: (a) 4P, (b) 4P+M, (c) 3P, and (d) 3P+M. Curves show the time evolution of the scaled energies $E_b/\omega_b$ and $E_c/\omega_c$ under left- and right-mode driving. Here, $\omega_b$ and $\omega_c$ denote the frequencies of the QB and charger modes, respectively: under left-mode driving, $\omega_b=\omega_2$ and $\omega_c=\omega_1$, while under right-mode driving, $\omega_b=\omega_1$ and $\omega_c=\omega_2$. Red and yellow triangles denote the remote battery and driven-mode energies under left-mode driving, $E_b^{\mathrm{L,nr}}$ and
$E_c^{\mathrm{L,nr}}$, respectively. Green and blue squares show the corresponding energies under right-mode driving; here $E_b^{\mathrm{R,nr}}$ denotes the energy stored in the remote receiver (the left mode $a_1$), while $E_c^{\mathrm{R,nr}}$ is the energy in the directly driven mode on the right ($a_2$). In all setups the nonreciprocal condition is satisfied and the system is tuned to effective resonance, $\Delta_1^{\mathrm{eff}}=\Delta_2^{\mathrm{eff}}=0$. Parameters are $\gamma=0.01 \omega_0$,
$\kappa_1=\kappa_2=0.01 \omega_0$, and $\Omega=0.15 \omega_0$, where $\omega_0\approx\omega_1\approx\omega_2$. We use equal waveguide couplings $\gamma_j=\gamma$, and the phase choices are given in Table~\ref{tab:nonreciprocal_points}.}
\end{figure}
In Fig.~\ref{fig:time_dynamics}, the scaled energies of the charger $E_c^{\mathrm{L,nr}}/\omega_1$ and the QB $E_b^{\mathrm{L,nr}}/\omega_2$ are plotted versus the scaled time $\gamma t$. In all configurations the energy in the charger increases rapidly at early times, while the QB's energy grows more gradually and reaches its steady state only after the charger has substantially saturated. This behavior directly reflects the cascaded structure of the dynamics imposed by the nonreciprocal condition $g_{\leftarrow}=0$, under which the driven mode acts as a source that feeds the remote mode without receiving dynamical feedback from it. Consequently, the charger energy builds up first and subsequently charges the QB.

Apart from this general trend, Fig.~\ref{fig:time_dynamics} also shows clear differences between left- and right-mode-driven cases. For driving applied on the left mode, the QB is charged in all four setups, as shown by the growth of $E_b^{\mathrm {L,nr}}$ (red curves with triangles). Its steady-state value under the nonreciprocal condition is
\begin{equation}
E_{\mathrm b}^{\mathrm{L,nr}}(\infty)
=
\frac{16\,\omega_2\,|g_{\rightarrow}|^2\Omega^2}{\Lambda_1^2\Lambda_2^2},
\end{equation}
which follows from the long-time limit of Eq.~\eqref{eq:nonreciprocal_energies}. The mirror-terminated setups (4P+M and 3P+M) achieve much larger $E_b^{\mathrm L}$ than the corresponding open setups (4P and 3P). Physically, the mirror redirects the propagating field back to the emitters and enhances constructive interference, which increases the effective forward coupling and thus the stored battery energy. In particular, the two mirror-terminated setups give the same steady-state battery energy (see the red curves with triangles in Fig.~\ref{fig:time_dynamics}). This does not result from identical effective parameters. Rather, at the nonreciprocal phase points in Table~\ref{tab:nonreciprocal_points}, the 4P+M and 3P+M setups have the same value of $|g_{\rightarrow}|/\Lambda_1$ and the same $\Lambda_2$, which yields the same $E_b^{\mathrm L}(\infty)$ according to Eq.~\eqref{eq:nonreciprocal_energies}, even though $|g_{\rightarrow}|$ and $\Lambda_1$ are individually different. A comparison between four-point and three-point setups shows that, the three-point setups perform as well as, or better than, the four-point ones. This suggests that, under the nonreciprocal condition, the three-point setups exhibit a stronger tendency for energy to accumulate in the QB among the four setups. In addition, only 4P and 3P+M satisfy $E_b^{\mathrm L}>E_c^{\mathrm L}$, indicating more efficient transfer of the input energy into the QB.

When the driving is applied on the right mode, the remote QB is not charged in any setups, which is different from the left-mode-driving case, as shown by the green curves with squares in Fig.~\ref{fig:time_dynamics}. The expressions for $E_b^{\mathrm R}(t)$ and $E_c^{\mathrm R}(t)$ have a form similar to Eqs.~\eqref{eq:nb_t_t_general} and \eqref{eq:nc_t_general} but with $g_{\rightarrow}$ replaced by $g_{\leftarrow}$. Under the nonreciprocal condition $g_{\leftarrow}=0$, the energies in the charger and QB are given by
\begin{subequations}\label{eq:ER_NR}
\begin{align}
E_b^{\mathrm{R,nr}}(t) &= 0,
\label{eq:EbR_NR_t}\\
E_c^{\mathrm{R,nr}}(t)
&=
\frac{4\,\omega_2\,\Omega^2}{\Lambda_2^2}
(1-e^{-\Lambda_2 t/2})^2.
\label{eq:EcR_NR_t}
\end{align}
\end{subequations}
The vanishing of the right-driven battery energy in Eq.~\eqref{eq:EbR_NR_t}
is a direct consequence of the selected forward nonreciprocal condition,
rather than an artifact of asymmetric local parameters. Under right-mode
driving, the charger is the right cavity mode $a_2$, whereas the remote QB is
the left cavity mode $a_1$. Hence, the charging of QB requires the backward waveguide-mediated channel $a_2\rightarrow a_1$, whose effective coupling strength is $g_{\leftarrow}$. The imposed phase condition cancels this channel, $g_{\leftarrow}=0$, while retaining a finite forward coupling
$g_{\rightarrow}\neq0$. Consequently, all source terms that could populate
$a_1$ are removed from the right-driven dynamics. Therefore, the energy in the charger (right mode) cannot be transferred to the QB (left mode) under the forward nonreciprocal condition (also see the green curves with squares in Fig.~\ref{fig:time_dynamics}). In the long-time limit, the energy in the charger for the four setups approaches the same steady-state value,
\begin{equation}
E_{\mathrm{c}}^{\mathrm{R,nr}}(\infty)
=
\frac{4\,\omega_2\,\Omega^2}{\Lambda_2^2}.
\label{eq:EcR_ss_NR}
\end{equation}
Equation~\eqref{eq:EcR_ss_NR} indicates that $E_{\mathrm{c}}^{\mathrm{R,nr}}(\infty)$ depends only on the local linewidth $\Lambda_2$ and the driving amplitude $\Omega$. Since all the propagation phases and coupling phases of the four setups in Table~\ref{tab:nonreciprocal_points} give the same $\Lambda_2$, the charger in these setups has the identical steady-state energy.


\begin{figure}[t]
\centering
\includegraphics[width=\linewidth]{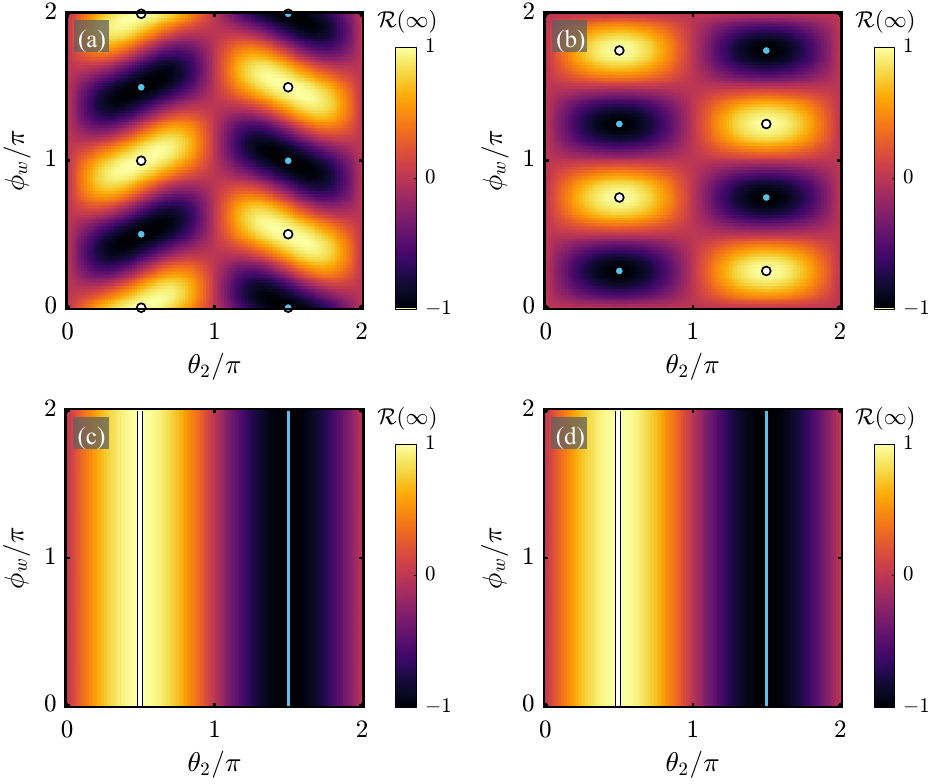}
\caption{\label{fig:R_maps}
Contour maps of the steady-state nonreciprocal ratio $\mathcal{R}(\infty)$ as a function of the QB coupling phase $\theta_2$ and the propagation phase $\phi_w$. The white markers indicate parameter values for which $\mathcal{R}(\infty)=1$, while the cyan-blue markers indicate parameter values for which $\mathcal{R}(\infty)=-1$. In panels (a) and (b), these conditions are marked by dots; in panels (c) and (d), they are marked by vertical lines. The setups correspond to (a) 4P, (b) 4P+M, (c) 3P, and (d) 3P+M. Other parameters used are the same as those in Fig.~\ref{fig:time_dynamics}.
}
\end{figure}

Furthermore, the underlying physics of reciprocity and nonreciprocity can be understood by examining the nonreciprocal ratio $\mathcal{R}(t)$ of the battery energies under left- and right-mode driving cases as 

\begin{equation}
\mathcal{R}(t)= \frac{E_b^{\mathrm{L}}(t)-E_b^{\mathrm{R}}(t)}
{E_b^{\mathrm{L}}(t)+E_b^{\mathrm{R}}(t)},
\label{eq:nonreciprocalr_ratio_defination}
\end{equation}
with $-1\leq \mathcal{R}(t)\leq 1$, provided that
$E_b^{\mathrm{L}}(t)+E_b^{\mathrm{R}}(t)\neq 0$. Here, $\mathcal{R}(t)=1$ corresponds to forward nonreciprocal charging, where energy flows unidirectionally from the left charger to the right battery ($E_b^{\mathrm{L}}>0$ and $E_b^{\mathrm{R}}=0$). Conversely, $\mathcal{R}(t)=-1$ corresponds to backward nonreciprocal charging, where energy flows from the right charger to the left battery ($E_b^{\mathrm{L}}=0$ and $E_b^{\mathrm{R}}>0$). The case $\mathcal{R}(t)=0$ indicates fully reciprocal charging. According to Eq.~\eqref{eq:nb_t_t_general}, the general steady-state energy of the QB for the left-mode-driven case is 
\begin{equation}\label{eq:Eb_ss_L_general}
E_{b}^{\mathrm L}(\infty)
=
\frac{16\,\omega_2\,|g_{\rightarrow}|^{2}\Omega^{2}}
{\big|4g_{\leftarrow}g_{\rightarrow}-\Lambda_1\Lambda_2\big|^{2}}.
\end{equation}
For the right-mode-driven case, we replace $g_{\leftarrow}$ with $g_{\rightarrow}$ to obtain 
\begin{equation}\label{eq:Eb_ss_R_general}
E_b^{\mathrm R}(\infty)
=
\frac{16\,\omega_1\,|g_{\leftarrow}|^{2}\Omega^{2}}
{\big|4g_{\leftarrow}g_{\rightarrow}-\Lambda_1\Lambda_2\big|^{2}}.
\end{equation}
With Eqs.~\eqref{eq:Eb_ss_L_general} and~\eqref{eq:Eb_ss_R_general}, the steady-state nonreciprocal ratio becomes

\begin{equation}
\mathcal{R}(\infty)
\approx
\frac{|g_{\rightarrow}|^{2}-|g_{\leftarrow}|^{2}}
{|g_{\rightarrow}|^{2}+|g_{\leftarrow}|^{2}},
\label{eq:Rss_general}
\end{equation}
where we have used the fact that the frequency mismatch between $\omega_1$ and $\omega_2$ is small and hence $\omega_1/\omega_2\approx 1$.
Equation~\eqref{eq:Rss_general} shows that, in the linear-driving case, the steady-state behavior is controlled only by the imbalance between the forward and backward effective couplings. 

To see the influence of quantum interference on steady-state nonreciprocal charging, we show $\mathcal{R}(\infty)$ as a function of the coupling phase $\theta_2$ and the propagation phase $\phi_w$ in Fig.~\ref{fig:R_maps}, where $\phi_w$ encodes the distance between the charger and the QB. The forward nonreciprocal points used in the time-dynamics simulations are marked by the white markers in Fig.~\ref{fig:R_maps}. For the four-point setups [Figs.~\ref{fig:R_maps}(a) and~\ref{fig:R_maps}(b)], the nonreciprocal response depends strongly on both $\theta_2$ and $\phi_w$. In the open 4P setup [Fig.~\ref{fig:R_maps}(a)], the regions with $\mathcal{R}(\infty)\simeq 1$ form diagonal lobe-like interference ridges, showing that forward directionality requires a simultaneous phase matching between the coupling phase and the propagation phase. The mirror-terminated 4P+M setup [Fig.~\ref{fig:R_maps}(b)] displays a more segmented pattern, with alternating forward and backward nonreciprocal domains generated by the additional reflected pathways. This reflects the richer multi-path interference enabled by two coupling points per emitter, which offers enhanced tunability at the cost of greater sensitivity to the charger--battery separation.

In contrast, the three-point setups [Figs.~\ref{fig:R_maps}(c)
and \ref{fig:R_maps}(d)] display vertical high-$\mathcal{R}(\infty)$
regions that are essentially independent of $\phi_w$. This distance
insensitivity can be seen directly from the directional couplings.
For equal waveguide coupling rates, the open 3P setup gives
\begin{align}
g_{\rightarrow}^{\mathrm{3P}}
&=
-\frac{\gamma}{2}e^{i\phi_w}
\left[1+e^{i(\phi_2-\theta_2)}\right],
\\
g_{\leftarrow}^{\mathrm{3P}}
&=
-\frac{\gamma}{2}e^{i\phi_w}
\left[1+e^{i(\phi_2+\theta_2)}\right],
\end{align}
whereas the mirror-terminated 3P+M setup gives
\begin{align}
g_{\rightarrow}^{\mathrm{3P+M}}
&=
-\gamma \cos\left(\tfrac{\phi_m}{2}\right)
e^{i(\phi_w+\phi_m/2)}
\left[1+e^{i(\phi_2-\theta_2)}\right],
\\
g_{\leftarrow}^{\mathrm{3P+M}}
&=
-\gamma \cos\left(\tfrac{\phi_m}{2}\right)
e^{i(\phi_w+\phi_m/2)}
\left[1+e^{i(\phi_2+\theta_2)}\right].
\end{align}
In both cases, the propagation phase $\phi_w$ enters only as a global phase factor. Therefore, it cancels from
$|g_{\rightarrow}|^2$ and $|g_{\leftarrow}|^2$, and hence from the steady-state nonreciprocal ratio. Using Eq.~\eqref{eq:Rss_general}, one obtains
\begin{equation}
\mathcal{R}_{\mathrm{3P}}(\infty)
=
\mathcal{R}_{\mathrm{3P+M}}(\infty)
\simeq
\frac{\sin\phi_2\,\sin\theta_2}
{1+\cos\phi_2\,\cos\theta_2}.
\label{eq:R_3P_3PM_simplified}
\end{equation}
For the phase choice
$\phi_2=\pi/2$, this expression reduces to
$\mathcal{R}(\infty)\simeq \sin\theta_2$, which explains the vertical
nonreciprocal bands in Figs.~\ref{fig:R_maps}(c) and
\ref{fig:R_maps}(d). Thus, the 3P and 3P+M configurations realize
distance-insensitive nonreciprocal charging because the charger--QB
separation affects only the overall phase of the directional coupling,
not its magnitude.


\begin{figure}[t]
\centering
\includegraphics[width=\linewidth]{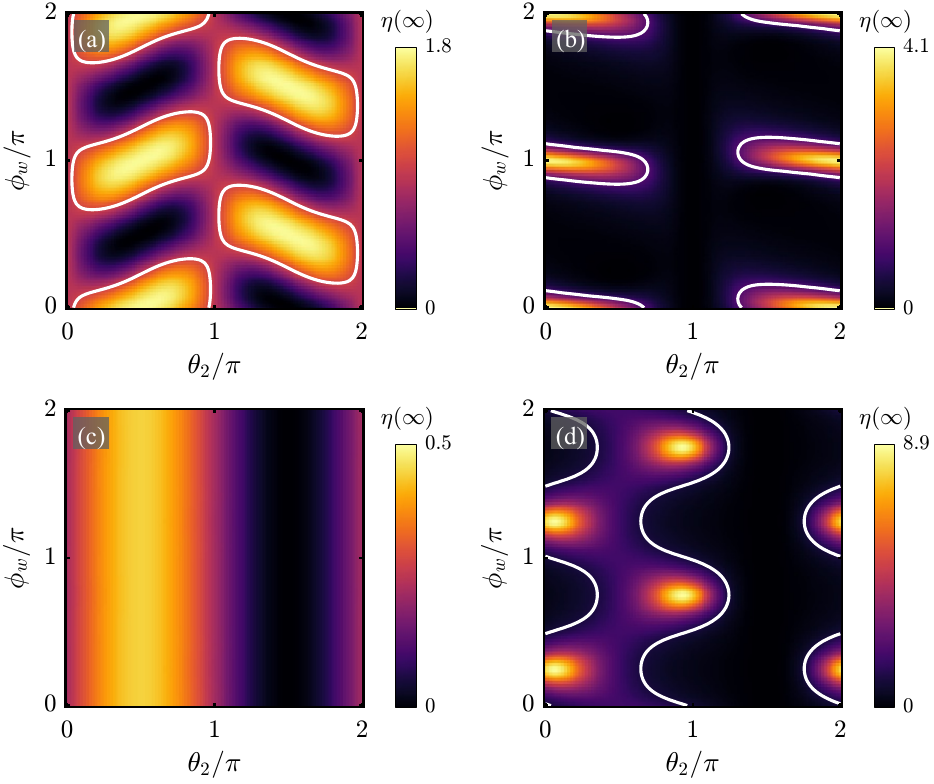}
\caption{\label{fig:eta_maps}
Relative storage ratio under left-mode driving. Color maps show the relative storage ratio $\eta(\infty)$ as a function of phase $\theta_2$ and $\phi_w$ for (a) 4P, (b) 4P+M, (c) 3P, and (d) 3P+M. The white contour marks $\eta=1$, separating battery-dominated storage ($\eta>1$) from charger-dominated storage ($\eta<1$). Other parameters used are the same as those in Fig.~\ref{fig:time_dynamics}.}

\end{figure}

While $\mathcal{R}$ quantifies the directional asymmetry between left- and right-mode driven cases, it does not specify how the injected energy is distributed between the charger and the QB. To characterize the
storage efficiency of the forward nonreciprocal charging process, we
focus on the left-mode-driven case, for which the selected phase condition
allows energy transfer from the charger $a_1$ to the remote QB $a_2$. We therefore introduce the relative storage ratio $\eta(t) = E_b^{\mathrm L}(t)/E_c^{\mathrm L}(t)$, with $\eta(t)>1$ corresponding to the battery-dominated storage, whereas $\eta(t)<1$ indicating the charger-dominated storage. Using Eqs.~\eqref{eq:nb_t_t_general} and \eqref{eq:nc_t_general}, the steady-state relative storage ratio is

\begin{equation}
\eta(\infty)= \frac{E_b^{\mathrm L}(\infty)}{E_c^{\mathrm L}(\infty)}
\approx
\frac{4\,|g_{\rightarrow}|^{2}}{\Lambda_2^{2}}.
\label{eq:eta_ss}
\end{equation}
Equation~\eqref{eq:eta_ss} shows that, under the left-mode-driven nonreciprocal operation, $\eta(\infty)$ is determined by the competition between the forward transfer strength $|g_{\rightarrow}|$ and the total linewidth $\Lambda_2$ of the QB. Therefore, nearly perfect nonreciprocity alone does not guarantee efficient remote storage: in addition to suppressing the backward coupling, one also needs to improve the ratio $|g_{\rightarrow}|/\Lambda_2$.

Figure~\ref{fig:eta_maps} shows the steady-state relative storage ratio for the four setups. The open 4P setup in Fig.~\ref{fig:eta_maps}(a) exhibits extended lobe-like regions with $\eta(\infty)>1$. These high-storage regions have a substantial overlap with the high-$\mathcal{R}(\infty)$ ridges in Fig.~\ref{fig:R_maps}(a), indicating that, in the open four-point setup, the interference condition that suppresses the backward channel can also enhance the forward transfer strength. By contrast, the mirror-terminated 4P+M setup in Fig.~\ref{fig:eta_maps}(b) is dominated by narrow high-$\eta$ regions embedded in broad $\eta(\infty)<1$ regions, showing that the mirror makes the storage ratio more phase selective. Although large values of $\eta(\infty)$ can be obtained, the high-$\eta$ regions do not systematically coincide with all perfect-nonreciprocal points. This indicates that the mirror-induced reflected pathways can decouple directionality from storage efficiency. The 3P setup in Fig.~\ref{fig:eta_maps}(c) remains entirely below unity, indicating that strong directional transport does not necessarily imply battery-dominated storage. In contrast, the 3P+M setup in Fig.~\ref{fig:eta_maps}(d) displays enhanced storage performance. The difference between the open and mirror-terminated three-point geometries can be made explicit by substituting the directional coupling and the QB linewidth into Eq.~\eqref{eq:eta_ss}. For equal waveguide couplings, $\gamma_1=\gamma_{21}=\gamma_{22}=\gamma$, the open 3P setup
gives
\begin{equation}
\eta_{\mathrm{3P}}(\infty)
=
\frac{
4\gamma^2
\cos^2\!\left[\left(\phi_2-\theta_2\right)/2\right]
}{
\left[
\kappa_2+2\gamma\left(1+\cos\phi_2\cos\theta_2\right)
\right]^2
}.
\label{eq:eta_3P_general}
\end{equation}
In the phase scans of Fig.~\ref{fig:eta_maps}, $\phi_2=\pi/2$ is fixed.
Equation~\eqref{eq:eta_3P_general} then reduces to
\begin{equation}
\eta_{\mathrm{3P}}(\infty)
=
\frac{
4\gamma^2
\cos^2\!\left(\pi/4-\theta_2/2\right)
}{
(2\gamma+\kappa_2)^2
},
\label{eq:eta_3P_scan}
\end{equation}
which is independent of $\phi_w$. This explains why Fig.~\ref{fig:eta_maps}(c)
shows vertical stripes and why the open 3P setup remains below
$\eta(\infty)=1$ for the parameters used here. For the mirror-terminated 3P+M setup, the forward coupling is enhanced by the reflected pathway, while the QB linewidth is also modified by mirror-induced self-interference. The corresponding relative storage ratio is
\begin{equation}
\eta_{\mathrm{3P+M}}(\infty)
=
\frac{
16\gamma^2
\cos^2\!\left(\phi_m/2\right)
\cos^2\!\left[\left(\phi_2-\theta_2\right)/2\right]
}{
\left(\Lambda_2^{\mathrm{3P+M}}\right)^2
},
\label{eq:eta_3PM_general}
\end{equation}
with $
\Lambda_2^{\mathrm{3P+M}}
=
\kappa_2
+
2\gamma
|
\cos\!\left(\phi_w+\phi_m/2\right)
+
e^{i\theta_2}
\cos\!\left(\phi_w+\phi_2+\phi_m/2\right)
|^2$.
For the phase choice used in Fig.~\ref{fig:eta_maps}(d), Eqs.~\eqref{eq:eta_3PM_general} reduces to
\begin{equation}
\eta_{\mathrm{3P+M}}(\infty)
=
\frac{
16\gamma^2
\cos^2\!\left(\pi/4-\theta_2/2\right)
}{
\left[
\kappa_2
+
2\gamma\left(1-\sin(2\phi_w)\cos\theta_2\right)
\right]^2
}.
\label{eq:eta_3PM_scan}
\end{equation}
Equation~\eqref{eq:eta_3PM_scan} shows that, in contrast to the open 3P
case, the storage ratio in the 3P+M setup depends sensitively on
$\phi_w$ through the effective QB linewidth. The mirror therefore does
not change the distance-insensitive form of the nonreciprocal ratio
$\mathcal{R}(\infty)$, but it strongly reshapes the storage efficiency
by modulating $\Lambda_2^{\mathrm{3P+M}}$. In regions where the
mirror-induced self-interference suppresses the QB linewidth, the ratio
$|g_{\rightarrow}|/\Lambda_2$ is enhanced, producing the large
$\eta(\infty)$ peaks observed in Fig.~\ref{fig:eta_maps}(d).

At the representative
nonreciprocal point $\phi_m=0$ and $\phi_2=\theta_2=\pi/2$, one obtains
$|g_{\rightarrow}^{\mathrm{3P}}|=\gamma$ but
$|g_{\rightarrow}^{\mathrm{3P+M}}|=2\gamma$, while both geometries have
the same QB linewidth $\Lambda_2=2\gamma+\kappa_2$. Equation~\eqref{eq:eta_ss}
therefore gives
\begin{subequations}
\label{eq:eta_nr_3p_3pm}
\begin{align}
\eta^{\rm nr}_{\mathrm{3P}}(\infty)
&=
\frac{4\gamma^2}{(2\gamma+\kappa_2)^2},
\label{eq:eta_nr_3p}
\\
\eta^{\rm nr}_{\mathrm{3P+M}}(\infty)
&=
\frac{16\gamma^2}{(2\gamma+\kappa_2)^2}.
\label{eq:eta_nr_3pm}
\end{align}
\end{subequations}
Hence, the mirror preserves the distance-insensitive directionality of the three-point setup while enhancing the forward transfer amplitude without increasing the effective QB linewidth at this phase point. This converts the nonreciprocal channel into a battery-dominated storage channel. A comparison between the nonreciprocal points marked in Fig.~\ref{fig:R_maps} and the storage-ratio maps in Fig.~\ref{fig:eta_maps} further highlights that nonreciprocity and storage efficiency are governed by distinct interference requirements. Although all these representative points satisfy $\mathcal{R}(\infty)=1$, they do not all lie in battery-dominated regions. Specifically, the 4P and 3P+M setups give $\eta^{\mathrm{nr}}(\infty)=16/9\simeq1.78$, whereas the 4P+M and 3P setups give
$\eta^{\mathrm{nr}}(\infty)=4/9\simeq0.44$. Therefore, perfect nonreciprocity realizes unidirectional charging, but it is not by itself
sufficient to ensure $\eta(\infty)>1$.



\section{Remote nonreciprocal charging in the quadratic-driving case}
\label{subsec:Quadratic}
In this section, we study the nonreciprocal charging when the charger is applied with a quadratic driving, with the drive Hamiltonian $H_d^{(\mathrm{quad})}$ given by Eq.~\eqref{eq:Hd_quad}. In the linear-driving case, the vacuum evolves into a coherent state, the normally ordered moments factorize, and the
stored energy is fully extractable. In contrast, the quadratic drive acts as a two-photon parametric pump, generating anomalous correlators that generally make the QB state non-passive. This makes the ergotropy a performance metric distinct from the stored energy. We therefore analyze, in
addition to directionality, the stability condition, second-moment dynamics, stored energy, and extractable work. Replacing $H_d$ with $H_d^{(\mathrm{quad})}$ in the quantum master equation~\eqref{eq:master}, the time evolution of the first moments in the left-mode-driven case are given by
\begin{subequations}\label{eq:quad_first}
\begin{align}
\dot{\langle a_1\rangle}
&=-\left(i\Delta_{1}^{\mathrm{eff}}+\frac{\Lambda_{1}}{2}\right)\langle a_1\rangle
+g_{\leftarrow}\langle a_2\rangle
-i\Omega \langle a_1^{\dagger}\rangle,
\label{eq:quad_first_c}\\
\dot{\langle a_2\rangle}
&=-\left(i\Delta_{2}^{\mathrm{eff}}+\frac{\Lambda_{2}}{2}\right)\langle a_2\rangle
+g_{\rightarrow}\langle a_1\rangle.
\label{eq:quad_first_b}
\end{align}
\end{subequations}
The right-mode-driven case is obtained by removing the quadratic-driving term
$-i\Omega \langle a_1^\dagger\rangle$ from Eq.~\eqref{eq:quad_first_c}
and adding the term $-i\Omega \langle a_2^\dagger\rangle$ to Eq.~\eqref{eq:quad_first_b}. From Eq.~\eqref{eq:quad_first}, we see that the directional behavior is still determined by $g_{\rightarrow}$ and $g_{\leftarrow}$, as in the linear-driving case. The crucial difference lies in the structure of the first-moment equations. For linear driving [Eq.~\eqref{eq:EoMset}], the equations are inhomogeneous due to the constant driving term $-i\Omega$, while the homogeneous part is intrinsically stable (its coefficient matrix has eigenvalues with negative real parts). Consequently, a unique steady state always exists, and no further stability analysis is needed. For quadratic driving, the first moments [Eq.~\eqref{eq:quad_first}] are homogeneous, meaning there is no constant term. Instead, the driving amplitude $\Omega$ appears inside the coefficient matrix, coupling $\langle a_1\rangle$ with its conjugate $\langle a_1^\dagger\rangle$. For a sufficiently large value of $\Omega$, the coefficient matrix may acquire eigenvalues with positive real parts, which renders the system unstable. Therefore, we need to determine the range of $\Omega$ for which a steady state exists. Writing
$\mathbf v=(\langle a_1\rangle,\langle a_2\rangle,\langle a_1^\dagger\rangle,\langle a_2^\dagger\rangle)^{T}$, the first moments in the quadratic-driving case can be written as a compact form

\begin{equation}
\dot{\mathbf v}=\mathbf{M}(\Omega)\,\mathbf v,
\label{eq:drift_first_moments}
\end{equation}
where the drift matrix reads
\begin{equation}
\mathbf{M}(\Omega)=
\begin{pmatrix}
A_1 & g_{\leftarrow} & -i\Omega & 0\\
g_{\rightarrow} & A_2 & 0 & 0\\
i\Omega & 0 & A_1^{*} & g_{\leftarrow}^{*}\\
0 & 0 & g_{\rightarrow}^{*} & A_2^{*}
\end{pmatrix},
\label{eq:M_quad_first}
\end{equation}
with $A_{j=1,2}=-(i\Delta_j^{\mathrm{eff}}+\Lambda_j/2)$. A steady state exists only when all eigenvalues $\lambda_j$ of $\mathbf M(\Omega)$ satisfy $
\max_j \text{Re}\left\{\lambda_j(\mathbf M)\right\}<0$. In the numerical simulations that follow, we restrict ourselves to parameters that satisfy this stability condition, thereby ensuring a steady state in the quadratic-driving case.

For vacuum initial conditions, $\mathbf v(0)=0$. Because Eq.~\eqref{eq:drift_first_moments} is homogeneous and contains no displacement-type source term, the unique solution is $\mathbf v(t)=0$ for all times. This does not imply the absence of charging. Instead, the quadratic drive injects energy through the second-order sector, as reflected by the inhomogeneous source term in the equation for $\langle a_1^2\rangle$ below. Consequently, the quadratic charging dynamics is fully determined by the second moments. Substituting $\langle a_2\rangle(t)=0$ into Eqs.~\eqref{eq:ergotropy_def}--\eqref{eq:Db_def}, we obtain the scaled ergotropy as

\begin{equation}
\frac{\mathcal{W}_b^\mathrm{L}(t)}{\omega_2}
=\langle a_2^\dagger a_2\rangle
-\frac{\sqrt{(1+2\langle a_2^\dagger a_2\rangle)^2
-4|\langle a_2^2\rangle|^2}-1}{2},
\label{eq:ergotropy_quad_simplified}
\end{equation}
which shows that the anomalous correlator $\langle a_2^2\rangle$ also contributes to the extractable work. For left-mode driving, Eq.~\eqref{eq:master} gives a closed linear system for the relevant second moments,

\begin{subequations}\label{eq:quad_second_set}

\begin{equation}\label{eq:quad_nc}
\begin{aligned}
\dot{\langle a_1^{\dagger}a_1\rangle}
&=-\Lambda_{1}\langle a_1^{\dagger}a_1\rangle
  +2\text{Re}\left\{g_{\leftarrow}^*\langle a_2^{\dagger}a_1\rangle\right\}\\
  &\qquad-2\Omega\text{Im}\left\{\langle a_1^{2}\rangle\right\},
\end{aligned}
\end{equation}

\begin{equation}\label{eq:quad_nb}
\begin{aligned}
\dot{\langle a_2^{\dagger}a_2\rangle}
&=-\Lambda_{2}\langle a_2^{\dagger}a_2\rangle
  +2\text{Re}\left\{g_{\rightarrow}\langle a_2^{\dagger}a_1\rangle\right\},
\end{aligned}
\end{equation}

\begin{equation}\label{eq:quad_cb}
\begin{aligned}
\dot{\langle a_1 a_2\rangle}
&=-\left[i\!\left(\Delta_{2}^{\mathrm{eff}}+\Delta_{1}^{\mathrm{eff}}\right)
+\tfrac{\Lambda_{2}+\Lambda_{1}}{2}\right]\langle a_1 a_2\rangle \\
&\quad +g_{\leftarrow}\langle a_2^{2}\rangle
+g_{\rightarrow}\langle a_1^{2}\rangle
-i\Omega\langle a_2^{\dagger}a_1\rangle^{*},
\end{aligned}
\end{equation}

\begin{equation}\label{eq:quad_bdc}
\begin{aligned}
\dot{\langle a_2^{\dagger}a_1\rangle}
&=\left[i\!\left(\Delta_{2}^{\mathrm{eff}}-\Delta_{1}^{\mathrm{eff}}\right)
-\tfrac{\Lambda_{2}+\Lambda_{1}}{2}\right]\langle a_2^{\dagger}a_1\rangle \\
&\quad +g_{\leftarrow}\langle a_2^{\dagger}a_2\rangle
+g_{\rightarrow}^*\langle a_1^{\dagger}a_1\rangle- i\Omega\langle a_1 a_2\rangle^{*},
\end{aligned}
\end{equation}

\begin{equation}\label{eq:quad_c2}
\begin{aligned}
\dot{\langle a_1^2\rangle}
&=-\left(2i\Delta_{1}^{\mathrm{eff}}+\Lambda_{1}\right)\langle a_1^{2}\rangle
+2\,g_{\leftarrow}\langle a_1 a_2\rangle \\
&\quad -2i\Omega\langle a_1^{\dagger}a_1\rangle
-i\Omega,
\end{aligned}
\end{equation}

\begin{equation}\label{eq:quad_b2}
\begin{aligned}
\dot{\langle a_2^2 \rangle}
&=-\left(2i\Delta_{2}^{\mathrm{eff}}+\Lambda_{2}\right)\langle a_2^{2}\rangle
+2\,g_{\rightarrow}\langle a_1 a_2\rangle.
\end{aligned}
\end{equation}

\end{subequations}

Solving Eq.~\eqref{eq:quad_second_set} for left-mode quadratic driving gives the time evolution of the charger energy $E_c^\mathrm{L}(t)=\omega_1\langle a_1^\dagger a_1\rangle$, the battery energy $E_b^\mathrm{L}(t)=\omega_2\langle a_2^\dagger a_2\rangle$ and the battery ergotropy $\mathcal{W}_b^\mathrm{L}(t)$ via Eq.~\eqref{eq:ergotropy_quad_simplified}. Under the nonreciprocal condition, Eq.~\eqref{eq:quad_second_set} can be solved analytically. The resulting expressions are lengthy but straightforward to derive (see Appendix~\ref{app:quad_solution} for details).

\begin{figure}[t]
\centering
\includegraphics[width=\linewidth]{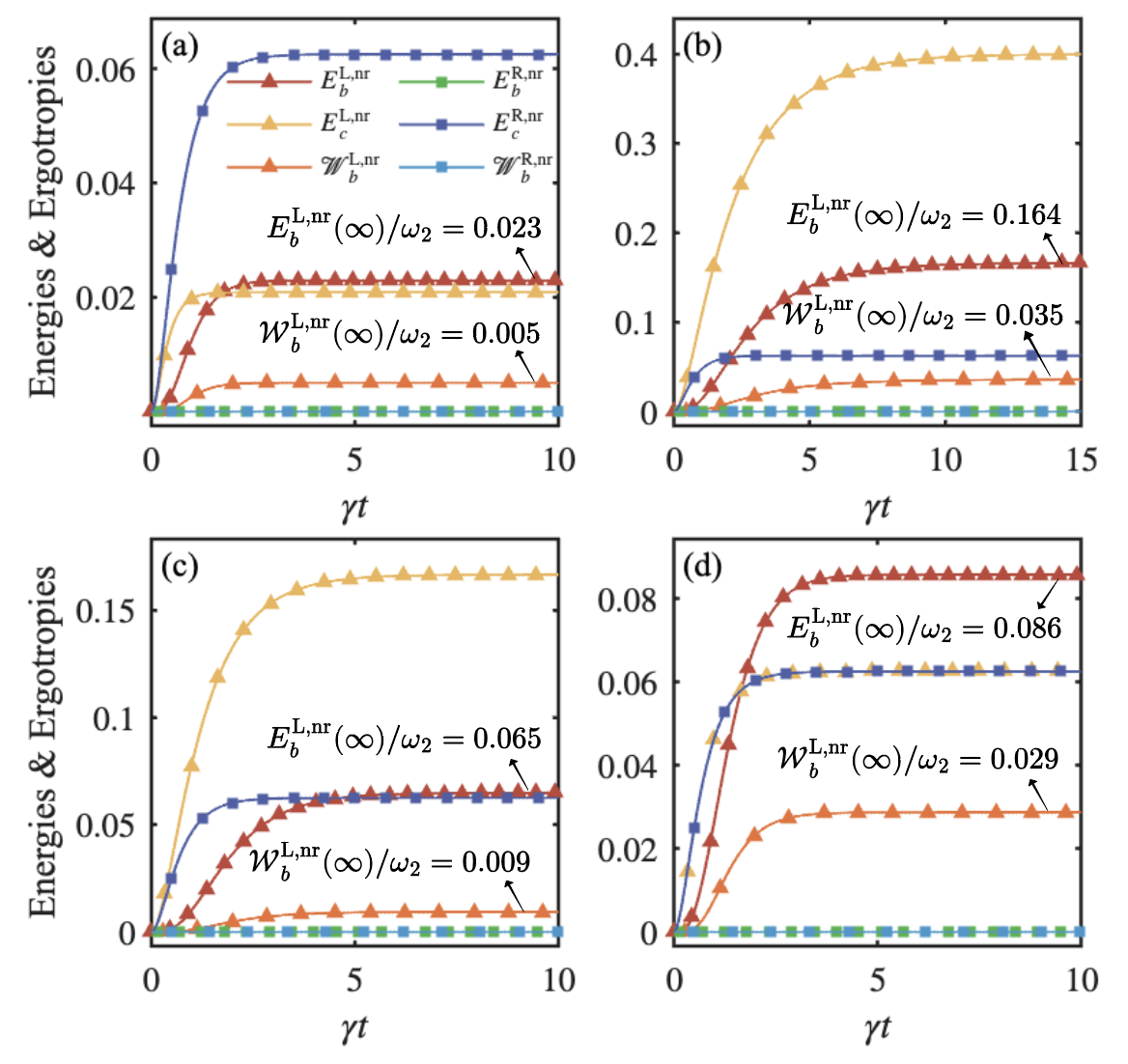}
\caption{\label{fig:time_dynamics_quad}
Remote charging dynamics under quadratic (two-photon) driving for the four configurations: (a) 4P, (b) 4P+M, (c) 3P, and (d) 3P+M. Curves show the time evolution of the scaled energies $E_b/\omega_b$, $E_c/\omega_c$, and the battery ergotropy $\mathcal{W}_b/\omega_b$ under left- and right-mode driving. Here, $\omega_b$ and $\omega_c$ denote the frequencies of the QB and charger modes, respectively: under left-mode driving, $\omega_b=\omega_2$ and $\omega_c=\omega_1$, while under right-mode driving, $\omega_b=\omega_1$ and $\omega_c=\omega_2$. Red and yellow curves with triangles denote the QB and driven charger energies under left-mode driving, $E_b^{\mathrm{L,nr}}$ and $E_c^{\mathrm{L,nr}}$, respectively. The corresponding energies under right-mode driving are shown in green and violet curves with squares, namely $E_b^{\mathrm{R,nr}}$ and $E_c^{\mathrm{R,nr}}$. The orange and blue curves show the battery ergotropy under left- and right-mode driving, $\mathcal{W}_b^{\mathrm{L,nr}}$ and $\mathcal{W}_b^{\mathrm{R,nr}}$, respectively. In all setups the nonreciprocal condition is satisfied and the system is tuned to effective resonance, $\Delta_1^{\mathrm{eff}}=\Delta_2^{\mathrm{eff}}=0$. Parameters are $\gamma=0.01\omega_0$, $\kappa_1=\kappa_2=0.01\omega_0$, and $\Omega=0.005\omega_0$, where $\omega_0\approx\omega_1\approx\omega_2$. We use equal waveguide couplings $\gamma_j=\gamma$, and phase values used are given in Table~\ref{tab:nonreciprocal_points}.}
\end{figure}

Figure~\ref{fig:time_dynamics_quad} displays the time evolution of the scaled energies $E_b/\omega_b$ and the battery ergotropy $\mathcal{W}_b/\omega_b$ for the four configurations under the quadratic-driving case. We take the driving amplitude $\Omega/\omega_0=0.005$, which guarantees the existence of a steady state for all setups. The smaller absolute energy scale in Fig.~\ref{fig:time_dynamics_quad}, compared with the linear-driving dynamics in Fig.~\ref{fig:time_dynamics}, mainly originates from the different driving amplitudes used in the two cases. For linear driving we take $\Omega=0.15\omega_0$, while for quadratic
driving we use $\Omega=0.005\omega_0$, so as to keep
the system below the parametric-instability threshold specified by Eq.~\eqref{eq:M_quad_first}. Thus, the reduced energy magnitude is a
consequence of the stability-constrained pump strength, rather than a generic limitation of the quadratic-driving mechanism. From Fig.~\ref{fig:time_dynamics_quad} we see that, in all setups, the left-mode driving changes the QB's energy $E_b^{\mathrm{L,nr}}(t)$ and it approaches a finite steady-state value, whereas the right-mode driving leaves the QB empty (i.e., $E_b^{\mathrm{R,nr}}(t)=0$ and $\mathcal{W}_b^\mathrm{R,nr}(t)=0$). The charging sequence is similar to the linear-driving case, where the charger energy $E_c^{\mathrm{L,nr}}(t)$ rises rapidly at early times, while the QB energy $E_b^{\mathrm{L,nr}}(t)$ builds up more gradually and reaches its steady state later.

The steady-state expressions further explain this behavior. From Eqs.~\eqref{eq:quad_closed_time_app} in Appendix~\ref{app:quad_solution}, the steady-state values, $E_b^{\mathrm{L,nr}}(\infty)$ and $E_c^{\mathrm{L,nr}}(\infty)$, are given by
\begin{subequations}\label{eq:quad_ss_Eb_Ec}
\begin{align}
E_c^{\mathrm{L,nr}}(\infty)
&=\frac{2\omega_1\Omega^{2}}{\Lambda_{1}^{2}-4\Omega^{2}},
\label{eq:Ec_ss_quad}\\[1mm]
E_b^{\mathrm{L,nr}}(\infty)
&=\frac{4\omega_2|\Gamma_{12}|^{2}(\Lambda_{2}+2\Lambda_{1})}{\omega_1\Lambda_{2}\left[(\Lambda_{2}+\Lambda_{1})^{2}-4\Omega^{2}\right]}E_c^{\mathrm{L,nr}}(\infty).
\label{eq:Eb_ss_quad}
\end{align}
\end{subequations}
Equation~\eqref{eq:Ec_ss_quad} shows that the steady-state energy of the charger depends only on the effective linewidth $\Lambda_1$ and the driving amplitude $\Omega$, and increases as $\Lambda_1$ approaches the parametric threshold $2\Omega$. However, Eq.~\eqref{eq:Eb_ss_quad} shows that the battery energy further depends on the collective decay rate $\Gamma_{12}$ and the effective linewidth $\Lambda_2$. The stable charging therefore requires both efficient excitation generation in the charger and sufficiently weak damping during the energy transport. This interpretation is consistent with the charging dynamics in Fig.~\ref{fig:time_dynamics_quad}, where energy first accumulates in the charger and is subsequently transferred to the QB on a longer timescale.

\begin{figure}[t]
\centering
\includegraphics[width=\linewidth]{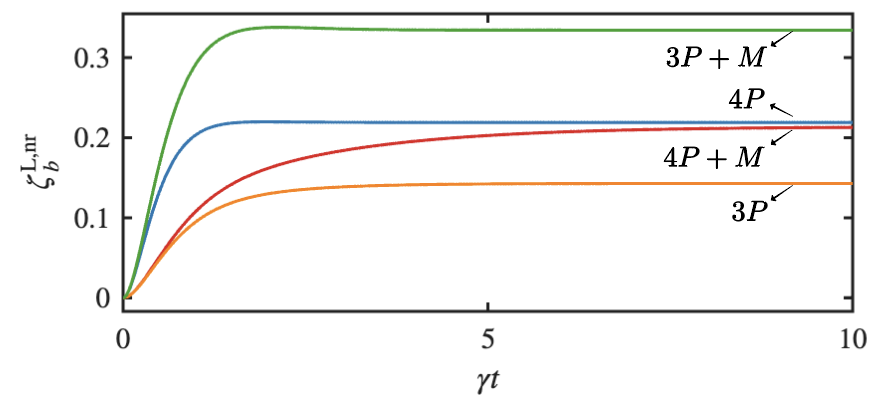}
\caption{\label{fig:Wfrac_quad}
Extractable fraction of the stored energy under quadratic driving. Curves show the ratio $\zeta_b^{\mathrm{L,nr}}(t)$ as a function of $\gamma t$ for the four configurations: 4P, 4P+M, 3P, and 3P+M. Other parameters used are the same as those in Fig.~\ref{fig:time_dynamics_quad}. }
\end{figure}


Comparing the four setups, the mirror-terminated configurations again exhibit enhanced energy accumulation. In particular, the 4P+M and 3P+M setups reach much larger steady-state energies in the QB than that in the open setups. The quadratic driving also changes the work-extraction properties of the QB. In Fig.~\ref{fig:time_dynamics_quad}, the ergotropy $\mathcal{W}_b^\mathrm{L,nr}(t)$ increases together with $E_b^\mathrm{L,nr}(t)$ but saturates at a lower value (see the orange and red curves with triangles). This is because linear driving produces a coherent state (a displaced vacuum) whose energy can be fully extracted by a reverse displacement. However, quadratic driving generates a squeezed state which contains correlations that prevent a simple displacement operation from extracting the full energy and hence only a fraction can be extracted as work. 

To quantify the extractable fraction of the stored energy, we introduce the ratio
\begin{equation}
\zeta_b^{\mathrm{L,nr}}(t)
=
\frac{\mathcal{W}_b^{\mathrm{L,nr}}(t)}
{E_b^{\mathrm{L,nr}}(t)} .
\label{eq:ergotropic_fraction}
\end{equation}
The time evolution of $\zeta_b^{\mathrm{L,nr}}(t)$ for the four setups is shown in Fig.~\ref{fig:Wfrac_quad}. It reveals a nontrivial interplay between coupling configurations and extractability. The 3P+M setup stores a larger fraction of its energy as work, while the 3P setup stores the smallest fraction. Thus, optimizing remote battery performance requires not only engineering nonreciprocity but also selecting a desired coupling configuration that favors high ergotropy.

\begin{figure}[t]
\centering
\includegraphics[width=\linewidth]{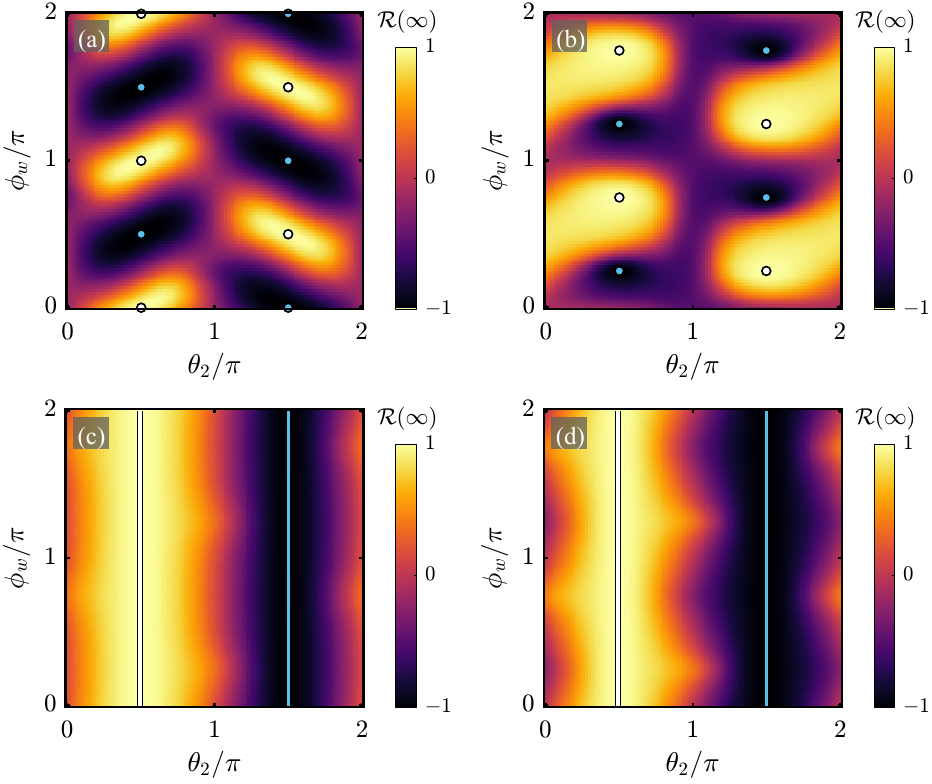}
\caption{\label{fig:R_maps_quad}
Contour maps of the steady-state nonreciprocal ratio $\mathcal{R}(\infty)$ as a function of the QB coupling phase $\theta_2$ and the propagation phase $\phi_w$. The white markers indicate parameter values for which $\mathcal{R}(\infty)=1$, while the cyan-blue markers indicate parameter values for which $\mathcal{R}(\infty)=-1$. In panels (a) and (b), these conditions are marked by dots; in panels (c) and (d), they are marked by vertical lines. The setups correspond to (a) 4P, (b) 4P+M, (c) 3P, and (d) 3P+M. Other parameters used are the same as those in Fig.~\ref{fig:time_dynamics_quad}.}
\end{figure}

Figure~\ref{fig:R_maps_quad} shows the steady-state nonreciprocal ratio $\mathcal{R}(\infty)$ under quadratic driving. The phase conditions for perfect forward and backward nonreciprocity, marked respectively by the white and cyan-blue dots and lines, coincide with those in the linear-driving case shown in Fig.~\ref{fig:R_maps}. This agreement is expected because the nonreciprocal condition is fixed by the cancellation of one directional waveguide-mediated coupling and is therefore independent of whether the system is driven linearly or quadratically. The surrounding phase landscape, however, is modified by the two-photon drive. In contrast to the linear-driving case, where Eq.~\eqref{eq:Rss_general} shows that $\mathcal{R}(\infty)$ is governed by the imbalance between $|g_{\rightarrow}|^2$ and $|g_{\leftarrow}|^2$, the quadratic response also involves anomalous second moments. Consequently, under quadratic driving the phase dependence of $\mathcal{R}(\infty)$ is no longer determined solely by the imbalance between $|g_{\rightarrow}|^2$ and $|g_{\leftarrow}|^2$. This drive-induced reshaping is most apparent in the mirror-terminated configurations.

In the open 4P setup [Fig.~\ref{fig:R_maps_quad}(a)], the quadratic-driving map closely reproduces the linear-driving pattern in Fig.~\ref{fig:R_maps}(a), with broad high-$\mathcal{R}(\infty)$ domains following nearly the same lobe-like structure. This indicates that, in this setup, the phase dependence of the directional response remains governed primarily by the underlying waveguide-mediated couplings and is only weakly affected by the change of driving mechanism. The mirror-terminated 4P+M setup [Fig.~\ref{fig:R_maps_quad}(b)] exhibits a more structured pattern, indicating that mirror-induced interference further reshapes the steady state in the quadratic-driving case. Furthermore, we observe nearly vertical high-$\mathcal{R}(\infty)$ bands in the three-point setups in Figs.~\ref{fig:R_maps_quad}(c) and~\ref{fig:R_maps_quad}(d), similar to the linear-driving case. Thus, their nonreciprocal response remains governed primarily by the QB coupling phase $\theta_2$, while the weak bending of the bands indicates a residual dependence on the propagation phase $\phi_w$ through the anomalous correlations. These results demonstrate that the quadratic- and linear-driving cases preserve the same perfect nonreciprocal points [the white and cyan-blue dots and lines in Figs.~\ref{fig:R_maps} and~\ref{fig:R_maps_quad}], but that quadratic driving reshapes the nearby phase dependence through phase-sensitive two-photon dynamics.


\begin{figure}[t]
\centering
\includegraphics[width=\linewidth]{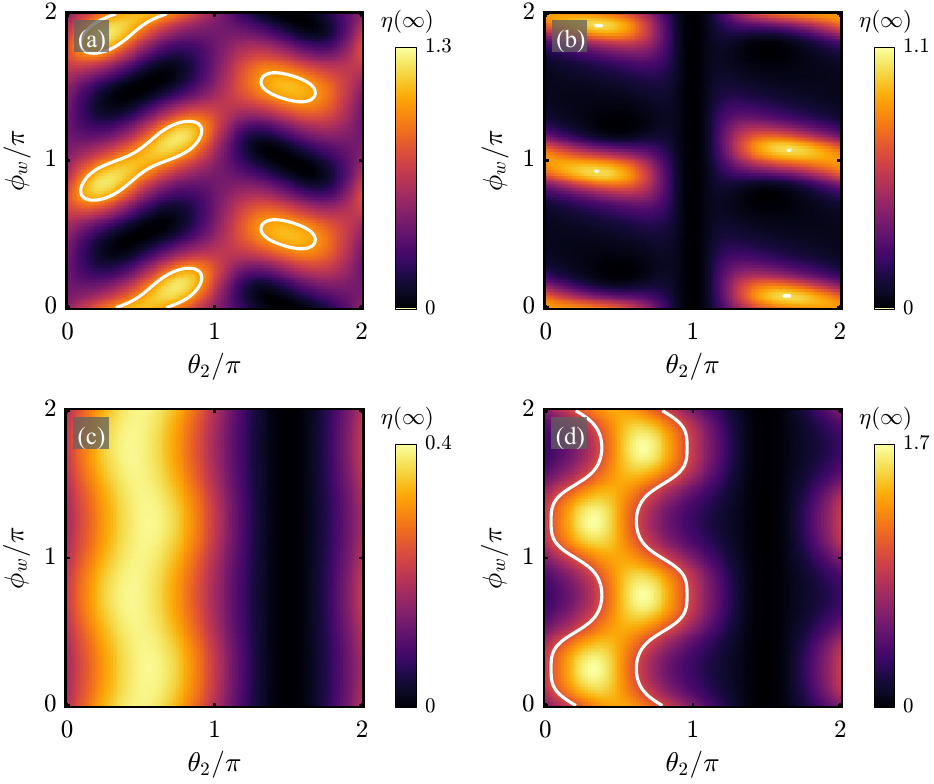}
\caption{\label{fig:Eb_over_Ec_quad}
Steady-state relative storage ratio $\eta(\infty)$ under quadratic driving as a function of the QB coupling phase $\theta_2$ and the propagation phase $\phi_w$. The four panels correspond to (a) 4P, (b) 4P+M, (c) 3P, and (d) 3P+M. The white contour marks $\eta(\infty)=1$. Other parameters used are the same as those in Fig.~\ref{fig:time_dynamics_quad}. }
\end{figure}

For the quadratic-driving case, the steady-state relative storage ratio $\eta(\infty)$ versus $\theta_2$ and $\phi_w$ for the four configurations is shown in Figure~\ref{fig:Eb_over_Ec_quad}. Compared with the linear-driving case in Fig.~\ref{fig:eta_maps}, the overall magnitude of $\eta(\infty)$ is reduced. In the open 4P setup [Fig.~\ref{fig:Eb_over_Ec_quad}(a)], the high-$\eta(\infty)$ regions retain a lobe-like structure and partially overlap with the high-$\mathcal{R}(\infty)$ domains in Fig.~\ref{fig:R_maps_quad}(a). The mirror-terminated 4P+M setup [Fig.~\ref{fig:Eb_over_Ec_quad}(b)] shows only localized regions with $\eta(\infty)>1$, confirming that mirror-terminated interference makes the storage efficiency more phase selective. The open 3P setup [Fig.~\ref{fig:Eb_over_Ec_quad}(c)] remains below $\eta(\infty)=1$ over the entire plotted region, even though its nonreciprocal ratio can approach unity in Fig.~\ref{fig:R_maps_quad}(c). By contrast, the mirror-terminated 3P+M setup [Fig.~\ref{fig:Eb_over_Ec_quad}(d)] supports broad battery-dominated regions with $\eta(\infty)>1$, although these regions acquire a visible $\phi_w$ modulation.


\section{Conclusion}
\label{sec:conclusion_outlook}

In conclusion, we have proposed a scheme to realize a waveguide-mediated remote quantum battery, in which a driven charger and a spatially separated battery are coupled through a one-dimensional waveguide. Via engineering the propagation and coupling phases, the effective coherent and dissipative couplings can be tuned to a nonreciprocal working point, yielding cascaded-like directional charging without any direct local charger--battery interaction. Our results show that the charging performance depends not only on directionality, as quantified by the nonreciprocal ratio, but also on how the injected energy is partitioned between the driven and remote modes, as captured by the relative storage ratio. The systematic comparison of the four coupling configurations further demonstrates that mirror termination and reduced coupling complexity can significantly reshape both the charging performance and the robustness of the operating phase window. Under quadratic driving, the generated squeezing correlations make the battery state non-passive, so that ergotropy becomes a distinct performance metric beyond the stored energy. These results identify giant-emitter waveguide platforms as a promising route toward programmable remote quantum-energy transfer and multi-node quantum-battery networks.


\appendix
\section{Derivation of the quantum master equations from the SLH formalism}
\label{appendixA}

In this appendix, we derive the quantum master equations in Eq.~\eqref{eq:master} for the four coupling configurations in Fig.~\ref{fig:schematic4p} by using the SLH formalism~\cite{GoughJames2009CMP,GoughJames2009IEEE,ZhangJames2012CSB,CombesKerckhoffSarovar2017}. In this approach, each element of the network is represented by an SLH triplet
\begin{equation}
G=(S,L,H),
\end{equation}
where $S$ is the scattering matrix, $L$ is the coupling operator to the propagating field, and $H$ is the internal Hamiltonian.

To combine different elements in the network, we use the series product. For two components
$G_{1}=(S_{1},L_{1},H_{1})$ and $G_{2}=(S_{2},L_{2},H_{2})$, connected such that the output of $G_{1}$ feeds into the input of $G_{2}$, the resulting triplet is
\begin{equation}
\label{eq:series_product_appendix}
\begin{aligned}
G_{2}\triangleleft G_{1}
&=
(
S_{2}S_{1},\;
L_{2}+S_{2}L_{1}, H_{res}).
\end{aligned}
\end{equation}
where the resulting Hamiltonian $H_{res}=H_{1}+H_{2}
+\frac{1}{2i}(L_{2}^{\dagger}S_{2}L_{1}-\mathrm{H.c.})$. 
For open-waveguide configurations, the right- and left-propagating channels are treated as independent and are combined through the concatenation product
\begin{equation}
G_{a} \boxplus G_{b}
=
\left(
\begin{bmatrix}
S_{a} & 0 \\
0 & S_{b}
\end{bmatrix},
\;
\begin{bmatrix}
L_{a}\\
L_{b}
\end{bmatrix},\;
H_{a}+H_{b}
\right).
\label{eq:concatenation_product_appendix}
\end{equation}

Propagation between two neighboring coupling points contributes only a phase shift. The SLH triplets for the propagation phases in Fig.~\ref{fig:schematic4p} are given by 

\begin{equation}
G_{\phi_{j=1,2,w,m}}=(e^{i\phi_j},0,0).
\label{eq:specific_phase_triplets_appendix}
\end{equation}

Each of these elements contributes only a phase and does not introduce any additional coupling operator or internal Hamiltonian. In principle, the propagation phases are frequency dependent, \(\phi=kd=\omega d/v\). One may therefore distinguish, for example, \(\phi_1=\omega_1 d/v\), \(\phi_2=\omega_2 d/v\), and \(\phi_w=\omega_0 d/v\). When the difference between \(\omega_1\) and \(\omega_2\) is very small (on the order of the waveguide coupling rate $\gamma$), we set \(\omega_0=(\omega_1+\omega_2)/2\) and use the approximation \(\omega_0\simeq\omega_1\simeq\omega_2\), and treat the propagation phases as defined with respect to a common reference frequency.

A local coupling point of a cavity mode to the waveguide is characterized by the coupling operator $L=\sqrt{\gamma/2}\,o$, where $o$ denotes the corresponding cavity-mode operator. In the SLH notation, the corresponding elementary component is written as
\begin{equation}
G=(1,L,H)=(1,\sqrt{\gamma/2}\,o,H).
\end{equation}
External driving terms are included directly in the internal Hamiltonian $H$. For each configuration, after constructing the total triplet
$G_{\mathrm{tot}}=(S_{\mathrm{tot}},L_{\mathrm{tot}},H_{\mathrm{tot}})$, we use the Lindblad dissipator $\mathcal{D}[o]\rho
=
o\rho o^{\dagger}
-\frac{1}{2}\{o^{\dagger}o,\rho\}$
to write out the quantum master equation that governs the reduced dynamics of the two cavity modes
\begin{equation}
\dot{\rho}
=
-i[H_{\mathrm{tot}},\rho]
+\mathcal{D}[L_{\mathrm{tot}}]\rho
+\kappa_{1}\mathcal{D}[a_1]\rho
+\kappa_{2}\mathcal{D}[a_2]\rho,
\label{eq:general_slh_master_appendix}
\end{equation}
where $L_{\mathrm{tot}}$ is a multichannel coupling vector,
$\mathcal{D}[L_{\mathrm{tot}}]\rho$ denotes the sum of the dissipators over all components of $L_{\mathrm{tot}}$. In addition to the waveguide-mediated decay contained in $L_{\mathrm{tot}}$, we also include intrinsic losses $\kappa_1$ and $\kappa_2$ of the two modes.

For open-waveguide configurations the system couples to two independent propagation channels corresponding to right- and left-propagating fields. By contrast, for mirror-terminated configurations the reflected field forms a single effective channel and the dynamics is described by a single coupling operator $L_{\mathrm{tot}}$.

By expanding Eq.~\eqref{eq:general_slh_master_appendix}, one can identify the effective detunings, coherent exchange interaction, individual decay rates, and collective decay rates that appear in Eq.~\eqref{eq:master} of the main text.

In what follows, we derive these coefficients separately for the open-waveguide and mirror-terminated configurations.

\subsection{Open-waveguide configurations}

For the open-waveguide configurations shown in Figs.~\ref{fig:schematic4p}(a) and (c), the right- and left-propagating fields are independent channels. The corresponding SLH models are therefore constructed separately and then combined through the concatenation product $G_{tot}=G_{R} \boxplus G_{L}$.

\subsubsection{Four-point setup (4P)}

In the four-point configuration in Fig.~\ref{fig:schematic4p}(a), both cavity modes are modeled as giant emitters with two coupling points each. The local SLH triplets for the right-propagating channel are
\begin{subequations}\label{eq:G_R_4P}
\begin{align}
G_{1,R}
&=(1,\sqrt{\gamma_{11}/2}\,a_1,\Delta_1 a_1^\dagger a_1 + H_d),
\label{eq:G1R}\\
G_{2,R}
&=(1,\sqrt{\gamma_{12}/2}\,e^{i\theta_1}a_1,0),
\label{eq:G2R}\\
G_{3,R}
&=(1,\sqrt{\gamma_{21}/2}\,a_2,\Delta_2 a_2^\dagger a_2),
\label{eq:G3R}\\
G_{4,R}
&=(1,\sqrt{\gamma_{22}/2}\,e^{i\theta_2}a_2,0),
\label{eq:G4R}
\end{align}
\end{subequations}
where $H_d$ is introduced in Eqs.~\eqref{eq:Hd_lin} and~\eqref{eq:Hd_quad}. Similarly, the left-propagating channel reads
\begin{subequations}\label{eq:G_L_4P}
\begin{align}
G_{1,L}
&=(1,\sqrt{\gamma_{11}/2}\,a_1,0),
\label{eq:G1L}\\
G_{2,L}
&=(1,\sqrt{\gamma_{12}/2}\,e^{i\theta_1}a_1,0),
\label{eq:G2L}\\
G_{3,L}
&=(1,\sqrt{\gamma_{21}/2}\,a_2,0),
\label{eq:G3L}\\
G_{4,L}
&=(1,\sqrt{\gamma_{22}/2}\,e^{i\theta_2}a_2,0),
\label{eq:G4L}
\end{align}
\end{subequations}
Note that for right-mode driving, the driving Hamiltonian $H_d$ is instead included in $G_{3,R}$, namely
$G_{3,R}=(1,\sqrt{\gamma_{21}/2}\,a_2,\Delta_2 a_2^\dagger a_2+H_d)$, while $G_{1,R}$ contains only the bare detuning term $\Delta_1 a_1^\dagger a_1$. The right-going chain is obtained by applying the series product along the propagation direction,
\begin{equation}\label{eq:GR}
G_R
=
G_{4,R}\triangleleft G_{\phi_2}
\triangleleft G_{3,R}\triangleleft G_{\phi_w}
\triangleleft G_{2,R}\triangleleft G_{\phi_1}
\triangleleft G_{1,R},
\end{equation}
while the left-going channel is
\begin{equation}\label{eq:GL}
G_L
=
G_{1,L}\triangleleft G_{\phi_1}
\triangleleft G_{2,L}\triangleleft G_{\phi_w}
\triangleleft G_{3,L}\triangleleft G_{\phi_2}
\triangleleft G_{4,L}.
\end{equation}

After repeated application of the series-product rule, the effective coupling operators can be written as
\begin{subequations}\label{eq:L_coeff}
\begin{align}
L_R = r_1 a_1+r_2 a_2, \quad L_L = l_1 a_1+l_2 a_2.
\label{eq:LR_coeff}
\end{align}
\end{subequations}
where the coefficients are
\begin{subequations}\label{eq:rl_coeff}
\begin{align}
r_1 &=
\sqrt{\tfrac{\gamma_{12}}{2}}e^{i(\theta_1+\phi_w+\phi_2)}
+\sqrt{\tfrac{\gamma_{11}}{2}}e^{i(\phi_1+\phi_w+\phi_2)},
\label{eq:r1_coeff}\\
r_2 &=
\sqrt{\tfrac{\gamma_{22}}{2}}e^{i\theta_2}
+\sqrt{\tfrac{\gamma_{21}}{2}}e^{i\phi_2},
\label{eq:r2_coeff}\\
l_1 &=
\sqrt{\tfrac{\gamma_{11}}{2}}
+\sqrt{\tfrac{\gamma_{12}}{2}}e^{i(\phi_1+\theta_1)},
\label{eq:l1_coeff}\\
l_2 &=
e^{i(\phi_1+\phi_w)}
\left(
\sqrt{\tfrac{\gamma_{21}}{2}}
+\sqrt{\tfrac{\gamma_{22}}{2}}e^{i(\phi_2+\theta_2)}
\right).
\label{eq:l2_coeff}
\end{align}
\end{subequations}

The total Hamiltonian takes the form
\begin{equation}
\label{eq:HamiltonianForm}
H_{\mathrm{tot}}
=
\sum_{j=1,2}\Delta_j^{\mathrm{eff}}a_j^\dagger a_j
+
\left(Ja_2^\dagger a_1+J^*a_1^\dagger a_2\right)
+
H_d .
\end{equation}

Using the master equation in Eq.~\eqref{eq:general_slh_master_appendix}, the individual and collective decay rates are obtained from
\begin{subequations}\label{eq:Gamma_coeff}
\begin{align}
\Gamma_{1} &= |r_1|^2 + |l_1|^2,
\label{eq:Gamma1_coeff}\\
\Gamma_{2} &= |r_2|^2 + |l_2|^2,
\label{eq:Gamma2_coeff}\\
\Gamma_{12} &= r_2 r_1^* + l_2 l_1^*.
\label{eq:Gamma12_coeff}
\end{align}
\end{subequations}

\subsubsection{Three-point setup (3P)}

In the three-point configuration, the left cavity mode $a_1$ is configured as a small emitter and couples to the waveguide at a single point, while the right cavity mode $a_2$ is configured as a two-point giant emitter. The SLH triplets for the right-propagating channel are
\begin{subequations}\label{eq:G_R_3P}
\begin{align}
G_{1,R}
&=(1,\sqrt{\gamma_1/2}\,a_1,\Delta_1 a_1^\dagger a_1 + H_d),
\label{eq:G1R_3P}\\
G_{2,R}
&=(1,\sqrt{\gamma_{21}/2}\,a_2,\Delta_2 a_2^\dagger a_2),
\label{eq:G2R_3P}\\
G_{3,R}
&=(1,\sqrt{\gamma_{22}/2}\,e^{i\theta_2}a_2,0).
\label{eq:G3R_3P}
\end{align}
\end{subequations}
with analogous expressions for the left-propagating channel. Following the same procedure as above, the resulting total Hamiltonian and the quantum master equation take the same form as in Eq.~\eqref{eq:HamiltonianForm} and Eq.~\eqref{eq:master}. 

\subsection{Mirror-terminated configurations}

For the mirror-terminated coupling configurations shown in Figs.~\ref{fig:schematic4p}(b) and (d), the mirror reflects the left-propagating field back into the system. As a result, the network can be represented as a single cascaded chain rather than two independent propagation channels. The total SLH triplet is therefore obtained directly through successive applications of the series product.

\subsubsection{Four-point setup (4P+M)}

In the case of the mirror-terminated four-point coupling configuration in Fig.~\ref{fig:schematic4p}(b), the total network is constructed by cascading all elements along the propagation direction,
\begin{align}
G_{\mathrm{tot}} =
G_{R}\triangleleft G_{\phi_m}\triangleleft G_{L},
\end{align}
where $G_{R}$ and $G_{L}$ are given in Eqs.~\eqref{eq:GR} and~\eqref{eq:GL}, respectively. Applying the series-product rule iteratively yields the effective SLH triplet $
G_{\mathrm{tot}}=(S_{\mathrm{tot}},L_{\mathrm{tot}},H_{\mathrm{tot}})$, 
where the total coupling operator can be written as $L_{\rm tot}=c_1 a_1+c_2 a_2$,
with
\begin{subequations}\label{eq:c_coeff}
\begin{align}
c_1
&=
e^{i(\phi_1+\phi_w)}
\Bigl[
\sqrt{\tfrac{\gamma_{12}}{2}}e^{i\theta_1}
\left(1+e^{i(\phi_m+2\phi_2)}\right)
\nonumber\\
&\qquad\qquad\qquad
+\sqrt{\tfrac{\gamma_{11}}{2}}e^{i\phi_2}
\left(1+e^{i\phi_m}\right)
\Bigr],
\label{eq:c1_coeff}\\[2mm]
c_2
&=
\sqrt{\tfrac{\gamma_{21}}{2}}e^{i\phi_1}
\left[
1+e^{i(\phi_m+2\phi_2+2\phi_w)}
\right]
\nonumber\\
&\qquad
+\sqrt{\tfrac{\gamma_{22}}{2}}e^{i\theta_2}
\left[
1+e^{i(2\phi_1+\phi_m+2\phi_2+2\phi_w)}
\right].
\label{eq:c2_coeff}
\end{align}
\end{subequations}

\subsubsection{Three-point setup (3P+M)}

In the mirror-terminated three-point coupling configuration in Fig.~\ref{fig:schematic4p} (d) the left mode $a_1$ couples to the waveguide at a single point, the SLH network is constructed by 
\begin{align}
G_{\mathrm{tot}} &=
G_{3,R}\triangleleft G_{\phi_2}
\triangleleft G_{2,R}\triangleleft G_{\phi_w}
\triangleleft G_{1,R}
\nonumber\\
&\quad\triangleleft
G_{\phi_m}\triangleleft G_{1,L}
\triangleleft G_{\phi_w}\triangleleft G_{2,L}
\triangleleft G_{\phi_2}\triangleleft G_{3,L}.
\end{align}

Repeated application of the series-product rule yields the total SLH triplet $G_{\mathrm{tot}}$, expanding the dissipator $\mathcal{D}[L_{\mathrm{tot}}]\rho$ yields the master equation reported in Eq.~\eqref{eq:master}, with the configuration-dependent coefficients listed in Appendix~\ref{app:coefficients}, Table~\ref{tab:coeff_all_4col}.


\clearpage

\onecolumngrid
\section{Coefficient expressions for the four coupling configurations}
\label{app:coefficients}
\suppressfloats[t]

For convenience, we collect here the explicit expressions of the waveguide-induced Lamb shifts, exchange interaction strength, and the individual and collective decay rates appearing in the master equation~\eqref{eq:master} for the four different coupling configurations discussed in the main text. The results are summarized in Table~\ref{tab:coeff_all_4col}.

\begin{table}[h]
\caption{\label{tab:coeff_all_4col}
Lamb shifts, exchange interaction strength, and individual and collective decay rates for all setups in Fig.~\ref{fig:schematic4p}. For the four-point configurations, we assume that all waveguide coupling rates are equal, $\gamma_{11}=\gamma_{12}=\gamma_{21}=\gamma_{22}=\gamma$. For the three-point configurations, we assume that all waveguide coupling rates are equal, $\gamma_{1}=\gamma_{21}=\gamma_{22}=\gamma$.}
\begin{ruledtabular}
\renewcommand{\arraystretch}{1.35}
\setlength{\extrarowheight}{3pt}
\setlength{\jot}{4pt}
\begin{tabular}{llll}
\parbox[c]{0.045\textwidth}{\centering \textbf{Coefficient}}
&
\parbox[c]{0.035\textwidth}{\centering \textbf{Setup}}
&
\multicolumn{2}{c}{Expressions} \\
& & \parbox[c]{0.18\textwidth}{\centering \textbf{Cavity mode $a_1$}}
& \parbox[c]{0.37\textwidth}{\centering \textbf{Cavity mode $a_2$}} \\
\hline

\multirow{4}{*}{\parbox{0.045\textwidth}{\raggedright Lamb shifts $\delta \omega_1, \,\delta \omega_2$}}
& \parbox[c]{0.035\textwidth}{\centering 4P}
& \parbox[t]{0.18\textwidth}{\raggedright
${\gamma}\sin\phi_{1}\cos\theta_{1}$}
& \parbox[t]{0.37\textwidth}{\raggedright
${\gamma}\sin\phi_{2}\cos\theta_{2}$}
\\

& \parbox[c]{0.035\textwidth}{\centering 4P+M}
& \parbox[t]{0.18\textwidth}{\raggedright
$\begin{aligned}[t]
&\gamma\cos\theta_{1}\!\left[\sin\phi_{2}
+\sin\!\left(\phi_{m}+\phi_{2}\right)\right] \\
&\,\,+\gamma/2\sin\phi_{m}
+\gamma/2\sin\!\left(\phi_{m}+2\phi_{2}\right)
\end{aligned}$}
& \parbox[t]{0.37\textwidth}{\raggedright
$\begin{aligned}[t]
&\gamma/2\,\sin\!\left(\phi_{m}+2\phi_{2}+2\phi_{w}\right)\\
&\,\,+\gamma/2\,\sin\!\left(\phi_{m}+2\phi_{2}+2\phi_{w}+2\phi_{1}\right) \\
&\,\,+\gamma\cos\theta_{2}\!\left[\sin\phi_{1}
+\,\sin\!\left(\phi_{m}+2\phi_{2}+2\phi_{w}+\phi_{1}\right)\right]
\end{aligned}$}
\\

& \parbox[c]{0.035\textwidth}{\centering 3P}
& \parbox[t]{0.18\textwidth}{\raggedright
$0$}
& \parbox[t]{0.37\textwidth}{\raggedright
${\gamma}\sin\phi_{2}\cos\theta_{2}$}
\\

& \parbox[c]{0.035\textwidth}{\centering 3P+M}
& \parbox[t]{0.18\textwidth}{\raggedright
$\frac{\gamma}{2}\sin\phi_{m}$}
& \parbox[t]{0.37\textwidth}{\raggedright
$\begin{aligned}[t]
&\gamma/2 \sin\!\left(\phi_{m}+2\phi_{w}\right)
+\gamma/2\sin\!\left(\phi_{m}+2\phi_{w}+2\phi_{2}\right) \\
&\,\,+\gamma\cos\theta_{2}\!\left[\sin\phi_{2}
+\sin\!\left(\phi_{m}+2\phi_{w}+\phi_{2}\right)\right]
\end{aligned}$}
\\

\hline

\multirow{4}{*}{\parbox{0.045\textwidth}{\raggedright Interaction $J$}}
& \parbox[c]{0.035\textwidth}{\centering 4P}
& \multicolumn{2}{l}{\parbox[t]{0.56\textwidth}{\raggedright
$\gamma/2 \!\left[\sin(\phi_{1}+\phi_{w})
+e^{-i\theta_{2}}\sin(\phi_{1}+\phi_{w}+\phi_{2})
+e^{i\theta_{1}}\sin\phi_{w}
+e^{i(\theta_{1}-\theta_{2})}\sin(\phi_{w}+\phi_{2})\right]$}}
\\

& \parbox[c]{0.035\textwidth}{\centering 4P+M}
& \multicolumn{2}{l}{\parbox[t]{0.56\textwidth}{\raggedright
\(\displaystyle
\begin{aligned}[t]
&
-i\,\gamma\,
e^{i\theta_{2}/2}
\left[
\cos(\phi_{m}/2)
+
e^{-i\theta_{1}}
\cos(\phi_{2}+\phi_{m}/2)
\right]
\\
&\;\times
\left[
e^{-i\left(\phi_{w}+\phi_{1}/2+\phi_{2}+\phi_{m}/2\right)}
\cos\left((\theta_{2}-\phi_{1})/2\right)-
e^{i\left(\phi_{w}+\phi_{1}/2+\phi_{2}+\phi_{m}/2\right)}
\cos\left((\theta_{2}+\phi_{1})/2\right)
\right]
\end{aligned}
\)
}}
\\

& \parbox[c]{0.035\textwidth}{\centering 3P}
& \multicolumn{2}{l}{\parbox[t]{0.56\textwidth}{\raggedright
$\gamma/2\!\left[\sin\phi_{w}+e^{-i\theta_{2}}\sin(\phi_{w}+\phi_{2})\right]$}}
\\

& \parbox[c]{0.035\textwidth}{\centering 3P+M}
& \multicolumn{2}{l}{\parbox[t]{0.56\textwidth}{\raggedright
$\gamma/2\,\!\left[\sin\phi_{w}+\sin(\phi_{m}+\phi_{w})
+e^{-i\theta_{2}}\!\left(\sin(\phi_{w}+\phi_{2})+\sin(\phi_{m}+\phi_{w}+\phi_{2})\right)\right]$}}
\\

\hline

\multirow{4}{*}{\parbox{0.045\textwidth}{\raggedright Waveguide-induced decay rate $\Gamma_{1}$}}
& \parbox[c]{0.035\textwidth}{\centering 4P}
& \multicolumn{2}{l}{\parbox[t]{0.56\textwidth}{\raggedright
$2\gamma\!\left[1+\cos\phi_{1}\cos\theta_{1}\right]$}}
\\

& \parbox[c]{0.035\textwidth}{\centering 4P+M}
& \multicolumn{2}{l}{\parbox[t]{0.56\textwidth}{\raggedright
\mbox{$
\gamma\left(1+\cos\phi_{m}\right)
+\gamma\!\left[1+\cos\!\left(\phi_{m}+2\phi_{2}\right)\right]
+2\gamma\cos\theta_{1}\!\left[\cos\phi_{2}+\cos\!\left(\phi_{m}+\phi_{2}\right)\right]
$}}}
\\

& \parbox[c]{0.035\textwidth}{\centering 3P}
& \multicolumn{2}{l}{\parbox[t]{0.56\textwidth}{\raggedright
$\gamma$}}
\\

& \parbox[c]{0.035\textwidth}{\centering 3P+M}
& \multicolumn{2}{l}{\parbox[t]{0.56\textwidth}{\raggedright
$\gamma\left[1+\cos\!\left(\phi_{m}\right)\right]$}}
\\

\hline

\multirow{4}{*}{\parbox{0.045\textwidth}{\raggedright Waveguide-induced decay rate $\Gamma_{2}$}}
& \parbox[c]{0.035\textwidth}{\centering 4P}
& \multicolumn{2}{l}{\parbox[t]{0.56\textwidth}{\raggedright
$2\gamma\!\left[1+\cos\phi_{2}\cos\theta_{2}\right]$}}
\\

& \parbox[c]{0.035\textwidth}{\centering 4P+M}
& \multicolumn{2}{l}{\parbox[t]{0.56\textwidth}{\raggedright
$\begin{aligned}[t]
&\gamma\!\left[1+\cos\!\left(\phi_{m}+2\phi_{2}+2\phi_{w}\right)\right]
+\gamma\!\left[1+\cos\!\left(2\phi_{1}+\phi_{m}+2\phi_{2}+2\phi_{w}\right)\right] \\
&\quad+2\gamma\cos\theta_{2}\!\left[\cos\phi_{1}
+\cos\!\left(\phi_{1}+\phi_{m}+2\phi_{2}+2\phi_{w}\right)\right]
\end{aligned}$}}
\\

& \parbox[c]{0.035\textwidth}{\centering 3P}
& \multicolumn{2}{l}{\parbox[t]{0.56\textwidth}{\raggedright
$2\gamma\!\left[1+\cos\phi_{2}\cos\theta_{2}\right]$}}
\\

& \parbox[c]{0.035\textwidth}{\centering 3P+M}
& \multicolumn{2}{l}{\parbox[t]{0.56\textwidth}{\raggedright
$\begin{aligned}[t]
&\gamma\!\left[1+\cos\!\left(\phi_{m}+2\phi_{w}\right)\right]
+\gamma\!\left[1+\cos\!\left(\phi_{m}+2\phi_{w}+2\phi_{2}\right)\right] \\
&\quad+2\gamma\cos\theta_{2}\!\left[\cos\phi_{2}
+\cos\!\left(\phi_{m}+2\phi_{w}+\phi_{2}\right)\right]
\end{aligned}$}}
\\

\hline

\multirow{4}{*}{\parbox{0.045\textwidth}{\raggedright Collective decay rate $\Gamma_{12}$}}
& \parbox[c]{0.035\textwidth}{\centering 4P}
& \multicolumn{2}{l}{\parbox[t]{0.56\textwidth}{\raggedright
$\gamma\!\left[
e^{i(\theta_{2}-\theta_{1})}\cos(\phi_{w}+\phi_{2})
+e^{i\theta_{2}}\cos(\phi_{1}+\phi_{w}+\phi_{2})
+e^{-i\theta_{1}}\cos\phi_{w}
+\cos(\phi_{1}+\phi_{w})
\right]$}}
\\

& \parbox[c]{0.035\textwidth}{\centering 4P+M}
& \multicolumn{2}{l}{\parbox[t]{0.56\textwidth}{\raggedright
\(\displaystyle
\begin{aligned}[t]
&
2\gamma
\left[
e^{i\theta_{1}}
\cos(\phi_{2}+\phi_{m}/2)
+
\cos(\phi_{m}/2)
\right]
\\
&\quad\times
\left[
\cos(\phi_{w}+\phi_{2}+\phi_{m}/2)
+
e^{-i\theta_{2}}
\cos(\phi_{1}+\phi_{w}+\phi_{2}+\phi_{m}/2)
\right]
\end{aligned}
\)
}}
\\

& \parbox[c]{0.035\textwidth}{\centering 3P}
& \multicolumn{2}{l}{\parbox[t]{0.56\textwidth}{\raggedright
$\gamma\!\left[\cos\phi_{w}+e^{-i\theta_{2}}\cos(\phi_{w}+\phi_{2})\right]$}}
\\

& \parbox[c]{0.035\textwidth}{\centering 3P+M}
& \multicolumn{2}{l}{\parbox[t]{0.56\textwidth}{\raggedright
$\gamma\!\left[\cos\phi_{w}+\cos(\phi_{m}+\phi_{w})
+e^{-i\theta_{2}}\!\left(\cos(\phi_{w}+\phi_{2})+\cos(\phi_{m}+\phi_{w}+\phi_{2})\right)\right]$}}
\\

\end{tabular}
\end{ruledtabular}
\end{table}

\twocolumngrid

\clearpage
\section{Analytical solutions for the quadratic-driving case under nonreciprocal condition}
\label{app:quad_solution}

Under the nonreciprocal condition, the second-moment equations \eqref{eq:quad_second_set} admit closed-form analytical solutions. The time-dependent expressions for the battery energy, charger energy, and anomalous correlators are given by
\begin{widetext}
\begin{subequations}\label{eq:quad_closed_time_app}
\begin{align}
E_b^\mathrm{L,nr}(t)
&=2\omega_2|\Gamma_{12}|^{2}\Omega\Bigg[
\frac{4\Omega(\Lambda_{2}+2\Lambda_{1})}
{\Lambda_{2}(\Lambda_{1}-2\Omega)(\Lambda_{1}+2\Omega)
(\Lambda_{2}+\Lambda_{1}-2\Omega)(\Lambda_{2}+\Lambda_{1}+2\Omega)}
\nonumber
+\frac{8\Omega(\Lambda_{2}-\Lambda_{1})e^{-\Lambda_{2}t}}
{\Lambda_{2}\left[(\Lambda_{2}-\Lambda_{1})^{2}-4\Omega^{2}\right]^{2}}
\nonumber\\
&\quad
-\frac{4e^{-\frac{t}{2}(\Lambda_{2}+\Lambda_{1}+2\Omega)}}
{(-\Lambda_{2}+\Lambda_{1}+2\Omega)^{2}(\Lambda_{2}+\Lambda_{1}+2\Omega)}
-\frac{e^{-t(\Lambda_{1}-2\Omega)}}
{(\Lambda_{1}-2\Omega)(\Lambda_{2}-\Lambda_{1}+2\Omega)^{2}}
\nonumber\\
&\quad
+\frac{4e^{-\frac{t}{2}(\Lambda_{2}+\Lambda_{1}-2\Omega)}}
{(\Lambda_{2}+\Lambda_{1}-2\Omega)(\Lambda_{2}-\Lambda_{1}+2\Omega)^{2}}
+\frac{e^{-t(\Lambda_{1}+2\Omega)}}
{(\Lambda_{1}+2\Omega)(-\Lambda_{2}+\Lambda_{1}+2\Omega)^{2}}
\Bigg],
\\
E_c^\mathrm{L,nr}(t)
&=-\omega_1 \frac{\Omega e^{-t(\Lambda_{1}+2\Omega)}
\left[\Lambda_{1}\left(e^{4t\Omega}-1\right)
+2\Omega\left(-2e^{t(\Lambda_{1}+2\Omega)}+e^{4t\Omega}+1\right)\right]}
{2\left(\Lambda_{1}^{2}-4\Omega^{2}\right)},
\\
\langle a_2^{2}(t)\rangle
&=2i|\Gamma_{12}|^{2}\Omega\Bigg[
-\frac{2\left[\Lambda_{1}(\Lambda_{2}+\Lambda_{1})+4\Omega^{2}\right]}
{\Lambda_{2}\big(\Lambda_{1}^{2}-4\Omega^{2}\big)
\left[(\Lambda_{2}+\Lambda_{1})^{2}-4\Omega^{2}\right]}
+\frac{2e^{-\Lambda_{2}t}\left[(\Lambda_{2}-\Lambda_{1})^{2}+4\Omega^{2}\right]}
{\Lambda_{2}\left[(\Lambda_{2}-\Lambda_{1})^{2}-4\Omega^{2}\right]^{2}}
\nonumber\\
&\quad
-\frac{4e^{-\frac{t}{2}(\Lambda_{2}+\Lambda_{1}+2\Omega)}}
{(-\Lambda_{2}+\Lambda_{1}+2\Omega)^{2}(\Lambda_{2}+\Lambda_{1}+2\Omega)}
+\frac{e^{-t(\Lambda_{1}-2\Omega)}}
{(\Lambda_{1}-2\Omega)(\Lambda_{2}-\Lambda_{1}+2\Omega)^{2}}
\nonumber\\
&\quad
-\frac{4e^{-\frac{t}{2}(\Lambda_{2}+\Lambda_{1}-2\Omega)}}
{(\Lambda_{2}+\Lambda_{1}-2\Omega)(\Lambda_{2}-\Lambda_{1}+2\Omega)^{2}}
+\frac{e^{-t(\Lambda_{1}+2\Omega)}}
{(\Lambda_{1}+2\Omega)(-\Lambda_{2}+\Lambda_{1}+2\Omega)^{2}}
\Bigg].
\end{align}
\end{subequations}
\end{widetext}

Equations~\eqref{eq:quad_closed_time_app} fully describe the time evolution of the system under quadratic driving at the nonreciprocal working point.

\end{document}